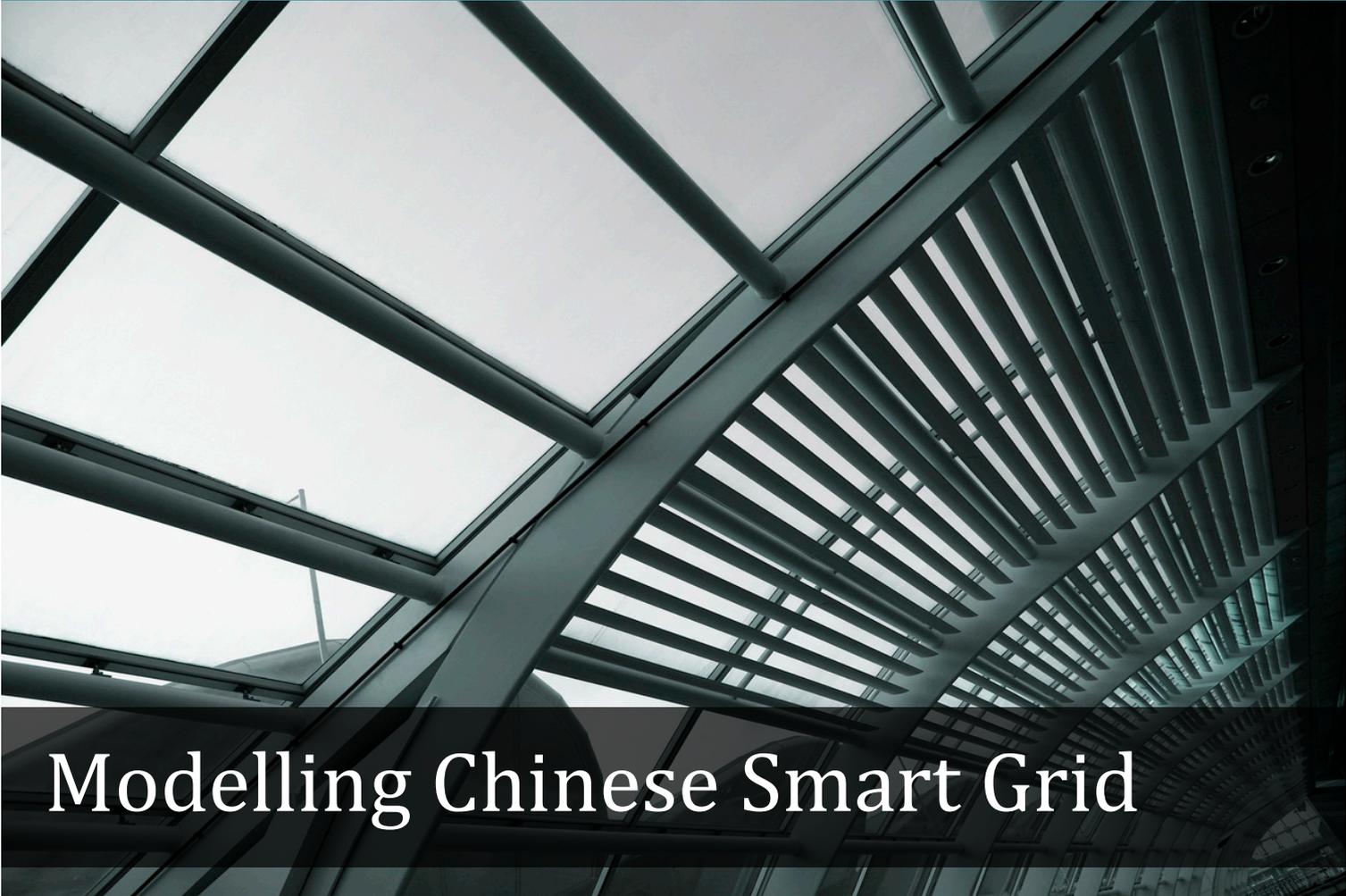

# Modelling Chinese Smart Grid

## A Stochastic Model Checking Case Study

Ender Yüksel  Huibiao Zhu  Heqing Huang
Hanne Riis Nielson
Flemming Nielson

Technical University of Denmark   East China Normal University   Wuxi SensingNet Industrialization Research Institute

# Table of Contents

# List of Tables

# List of Figures

# Foreword

In this document, we consider a specific Chinese Smart Grid implementation and try to address the verification problem for certain quantitative properties including performance and battery consumption. We employ stochastic model checking approach and present our modelling and analysis study using PRISM model checker.

# 1. Chinese Smart Grid

In this section, we define the cyber-physical system that we are considering.

## 1.1. Transmission Line and Towers

Electricity is transmitted through a power line, which is lifted by *tower*s connected in a linear arrangement. In a networking perspective, a (non-terminal) tower resembles to a node that has two connections: one to the predecessor tower and one to the successor tower.

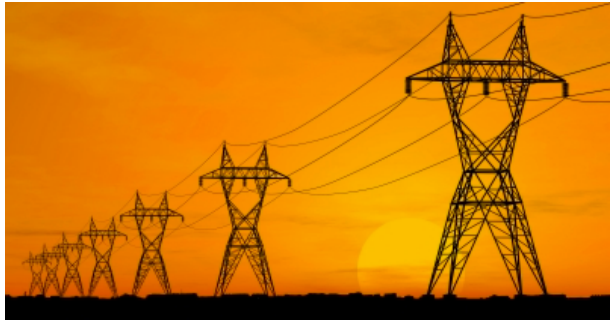

**Electricity transmission line.**

In a Smart Grid, we are also interested in the *data* transmission through the towers, which are accomplished by wireless sensor networks.

The towers are equipped with sensors such that each tower is data-wise connected to its closest two neighbours (solid lines) and less closer two other neighbours (dashed lines) as shown in Figure 1. The latter connection is considered as a back-up link and obviously it is more costly. A bone node needs to increase its RF power to 20dBm if it uses the back-up links.

There are 100 towers in total, and the distance between them is typically 200m. In other words, the total distance between two ends of the transmission line is about 20km.

A bone node is transmitting to the end that is cheaper to reach. So, normally the final destination of a packet is the closest end of the line, however one or more bone nodes in between is malfunctioning then the final destination could be the end that is physically more distant.

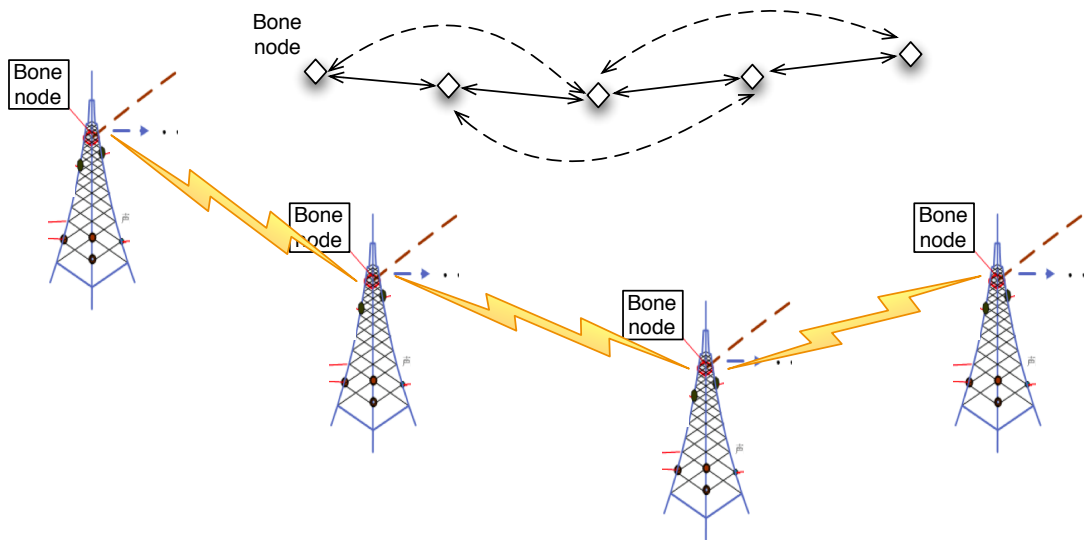

Figure 1: Wireless sensor network formed by bone nodes.

## 1.2. Wireless Sensor Network in a Tower

Each tower is equipped with a bone node and many different sensors (less than 50 sensors) as shown in Figure 2.

Sensors have different tasks and connected to the bone node in a star topology (the distance between sensors and the bone node is less than 50m).

Sensors are directly connected to the bone node, namely none of the sensors are functioning as a router.

There are eight types of sensors:
- distance measurement sensor
- voltage measurement sensor
- short circuit fault sensor
- vibration transducer (buried underground and hanging on tower)
- leaning degree measurement sensor
- line swing sensor
- wind direction sensor
- wind velocity sensor

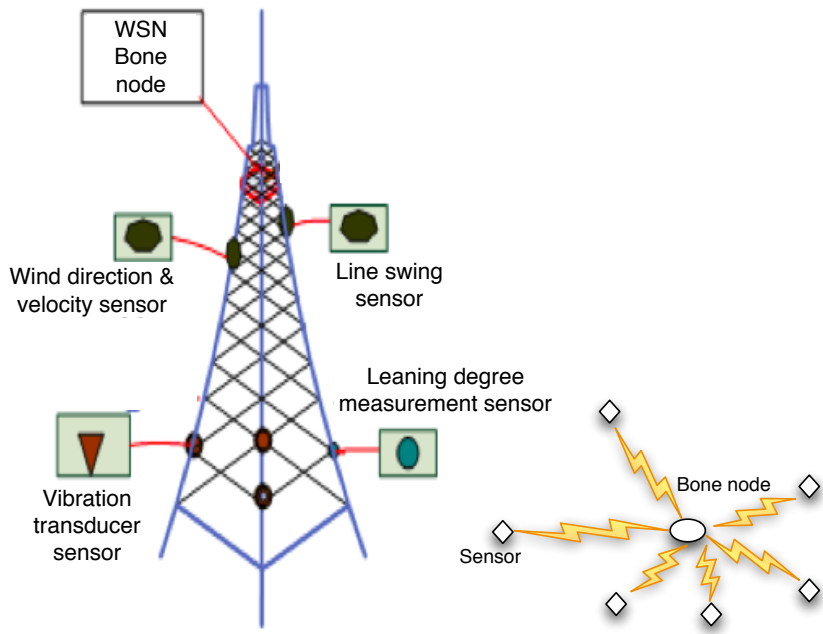

Figure 2: A tower equipped with sensors.

## 1.3. Energy Consumption of Sensors

The sensors have the same type of communication module, which makes their power consumption similar.

All sensors are battery operated and the capacity of a battery is about 1000mAh.

The energy consumption values for sensors and bone nodes are given in Table 1. Normally, TX(10dBm) is used for communication both between towers and within a tower. However, when using the backup-links, bone nodes increase their TX power from 10dBm to 20dBm.

Table 1: Energy consumption values.

|  | TX(3dBm) | TX(10dBm) | TX(20dBm) | RX | Sleep |
|---|---|---|---|---|---|
| **Sensor** | 40mA |  |  | 28mA | 5μA |
| **Bone node** |  | 120mA | 200mA | 18mA | 5μA |

Each sensor send one data packet per hour, and each packet is 40 Bytes in size.

The MAC protocol used in this system is based on the 802.15.4b MAC, and the data rate is 100kbps (at 470MHz).

## 1.4. The Goal

The power transmission line network, being a very critical part of the smart grid, should exhibit an appropriate robustness against failures. In order to achieve robustness, we should verify the low probability of errors, and good capabilities to recover from errors.

Sensors may fail over time because of physical reasons. In such case, failed sensor will be recovered and then it will continue serving. The goal is to ensure that the probability of a failure in the whole network is less than a pre-defined limit, e.g. 2%.

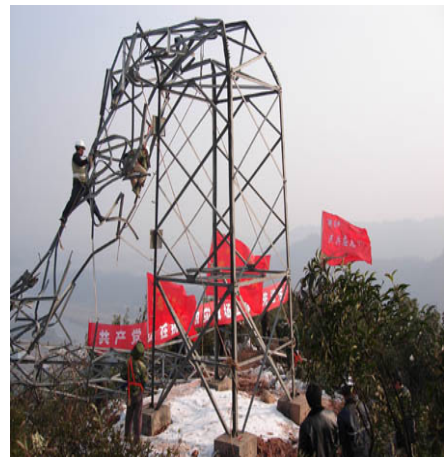

**Electricity workers repair the damaged power transmission tower in Yihuang County of South China's Jiangxi Province, Feb 12, 2008 (Xinhua Photo).**

## 2. Modelling

*We present a network topology in* Figure 3, *assuming:*
- *M Bone nodes,*
- *N sensor nodes in each tower,*
- *M+1 normal links between towers*
- *M backup links between towers*

*Abbreviations we used:*
- *BONE: Bone node on a tower*
- *S:    Sensor node on a tower*
- *L:    Normal link between sequential (neighbour) towers*
- *BL:   Backup link between non-neighbour towers*

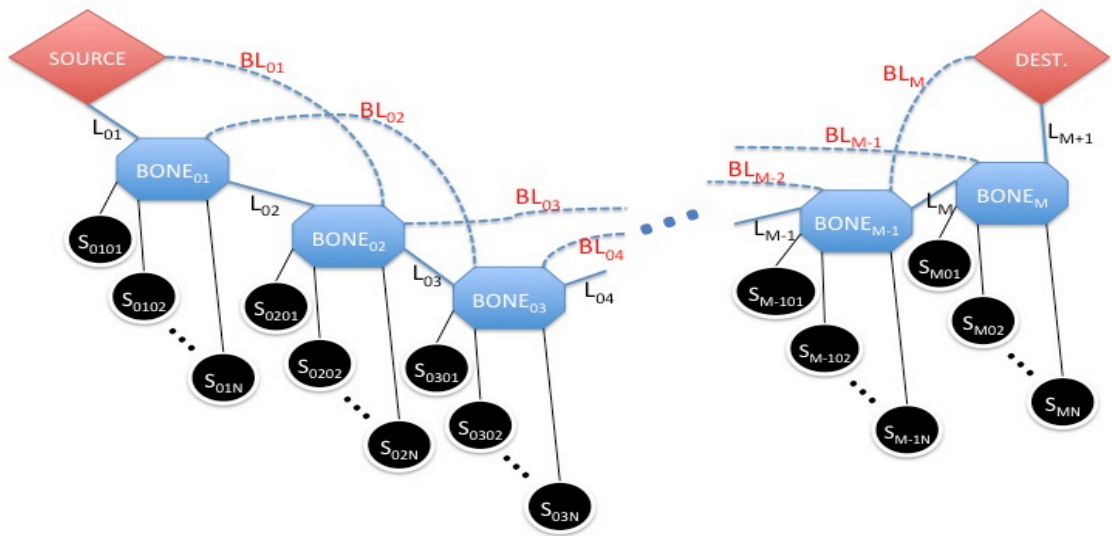

**Figure 3: The network topology.**

We assume that the philosophy of WSN is: *get the work done as quickly as possible and go to sleep.*

Sources of energy consumption:
1) **Sensing**, May be neglected.
2) **Processing**: May be neglected.[1]
3) **Communication**: Major consumer of energy

---

[1] The energy cost of transmitting 1 KB a distance of 100 meters is approximately 3 J (1 Joule = one watt per second) )(Power in Watts = Voltage in volts X Current in amperes). By contrast, a general-purpose processor with 100 MIPS/W power could execute 3 million instructions for the same amount of energy.
*Source: G. Pottie and W. Kaiser. Wireless integrated network sensors. In Communications of the ACM, volume 43, pages 551-8, may 2000.*

## 2.1. Limitations in Modelling

Modelling is tightly bound to the limitations about state-space. In order a formal model to be feasible for model checking, it should have a state space that is within certain limits. Below we present the situation with numerical values and examples.

- Total number of modules: #modules = (N+1)xM
    - M Bone nodes,
    - N sensor nodes in each tower,
    - In this real case: M=100, N=50, #modules=5100
- Example:
    - Assume that each module has 10 states.
        - Whole model: roughly $10^{5100}$ states.
        - Just a single tower model: roughly $10^{51}$ states.
    - Assume that each module has 2 states.
        - Whole model: roughly $2^{5100}$ states.
        - Just a single tower model: roughly $2^{51}$ (>$10^{15}$) states.
- If we are going to apply probabilistic model checking, then the total number of states should be much less
    - e.g. for PRISM less than $10^9$-$10^{10}$.

In conclusion, formal model of the whole network is not feasible for model checking without severe abstractions.

## 2.2. Formalization

### 2.2.1. Continuous-Time Markov Chain

A CTMC is a stochastic process with the property that every time it enters state $S_i$,
1. The amount of time the process spends in state $S_i$ before making a transition is exponentially distributed with some rate $R_i$, and
2. When the process leaves state $S_i$, it will next enter state $S_j$ with some probability $P_{ij}$ independent of the time spent at state $S_i$.

Continuous-Time Markov Chain (CTMC) models extend discrete-time Markov chains, where each transition corresponds to a discrete-time step. In CTMCs, transitions can occur in real time. Each transition is labelled with a **rate**, which defines the delay that occurs before the transition is taken. The delays are sampled from an exponential distribution that uses this rate as a parameter.

The probability of taking a transition from state *s* to state *s0* within *t* time units equals $1-e^{-R(s,s')\cdot t}$. In the case of the origin state *s* having more than one successor states s' such that **R**(s, s')>0, there exists a **race condition**. Namely, the transition taken will be the one, which is enabled first. Therefore, the probability

**P**(s, s') of moving from state s to state s' in a single transition is **R**(s, s')/E(s) where E(s) is called the **exit rate** and calculated as $\sum_{s' \in S} \mathbf{R}(s, s')$. Notice that, in the case of no outgoing transitions from s, **P**(s, s')=1 for s=s' and **P**(s, s')=0 for s!=s'.

The time spent in a state s before any transition happens (namely the **mean sojourn time**) is exponentially distributed with rate E(s).

## 2.3. The Road Map for Modelling

In the following three sections, we will model the Chinese Smart Grid in different perspectives. In each perspective, we will be using different abstractions. Our main trade-off will be between precision and scalability.

First, we will consider the inter-tower communication, abstracting intra-tower communications. In this case, we may focus more on the routing, and the usage of the expensive communication links.

Then, we will consider the intra-tower communication, abstracting inter-tower communication. In this case, we may focus more on the energy consumption of the sensor devices.

Finally, we will consider the whole network with heavy abstractions. In order to cover the whole network we will be sacrificing the precision on a single device. However, we will be able to see the big picture and analyze the whole network.

# 3. Transmission Line

In this section, we only consider the transmission line therefore just the bone nodes. We ignore the remaining sensor nodes that are deployed in electricity towers.

## 3.1. Routing

An important aspect in modelling the transmission line is the routing algorithm. In this case, we consider a routing scheme, which favors inexpensive transmission and tries to avoid the use of backup lines as much as possible. We explain and demonstrate this routing scheme in Appendix A. However, we should also formulate the routing in a way that can be formalized for model checking. To do so, we consider a simple topology of 10 towers, or 10 bone nodes. As shown in Figure 4, the bone nodes are connected with two different types of links: regular links shown with solid lines, and backup links showed with dashed lines. Terminal nodes are located in both ends, and they are only receiving data.

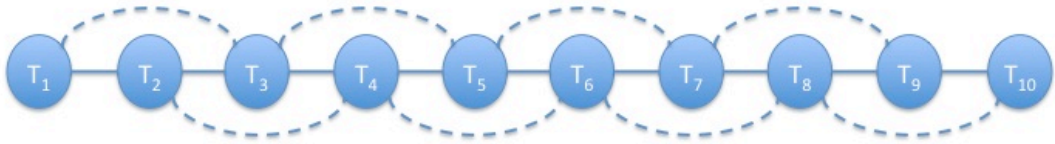

Figure 4: Example topology, composed of 10 bone nodes.

We studied the routing behaviour of each bone node, when the routing scheme in Appendix A is used. We start by mapping the links to the towers as we have listed in Table 2. In the first column, we have listed all the towers in the sample transmission line. Then we have two columns for mapping receive links, and two columns for mapping send links. First columns of receive and send links are the regular links, and the second columns are the backup links (denoted by XP). A link can be used at normal operation, such a link is denoted by a node pair e.g. "1-2". A link can also be used when rerouting, in this case it is followed by annotations in parenthesis e.g. "3-5 (!4&!6)||(!10&!4)". An annotation is simply a statement of the condition that should be satisfied in order to use that link.

Table 2: The mapping of the links to the towers.

| Tower | Receive Link | Expensive Receive Link | Send Link | Expensive Send Link |
|---|---|---|---|---|
| **T1** | 1-2 | 1-3 (!2&!4)\|\|(!10&!2) | | |
| **T2** | 2-3 | 2-4 (!3&!5)\|\|(!10&!3) | 1-2<br>2-3 (!1) | 2-4 (!1&!3) |
| **T3** | 2-3 (!1)<br>3-4 | 3-5 (!4&!6)\|\|(!10&!4) | 2-3<br>3-4 (!1\|\|!2) | 1-3 (!2&!4)\|\|(!10&!2)<br>3-5 (!1&!4) |
| **T4** | 3-4 (!1 \|\| !2)<br>4-5 | 2-4 (!1&!3)<br>4-6 (!10&!5) | 3-4<br>4-5 (!1\|\|!3) | 2-4 (!3&!5)\|\|(!10&!3)<br>4-6 (!1&!5) |
| **T5** | 4-5 (!1 \|\| !3)<br>5-6 (!7 \|\| !10) | 3-5 (!1&!4)<br>5-7 (!10&!6) | 4-5<br>5-6 (!1\|\|!4) | 3-5 (!4&!6)\|\|(!10&!4)<br>5-7 (!1&!6) |
| **T6** | 5-6 (!1 \|\| !4)<br>6-7 (!8 \|\| !10) | 4-6 (!1&!5)<br>6-8 (!10&!7) | 5-6 (!10\|\|!7)<br>6-7 | 4-6 (!10&!5)<br>6-8 (!1&!7)\|\|(!5&!7) |
| **T7** | 6-7<br>7-8 (!9 \|\| !10) | 5-7 (!1&!6)<br>7-9 (!10&!8) | 6-7 (!10\|\|!8)<br>7-8 | 5-7 (!10&!6)<br>7-9 (!1&!8)\|\|(!6&!8) |
| **T8** | 7-8<br>8-9 (!10) | 6-8 (!5&!7)\|\|(!1&!7) | 7-8 (!10\|\|!9)<br>8-9 | 6-8 (!10&!7)<br>8-10 (!1&!9)\|\|(!7&!9) |
| **T9** | 8-9 | 7-9 (!6&!8)\|\|(!1&!8) | 8-9 (!10)<br>9-10 | 7-9 (!10&!8) |
| **T10** | 9-10 | 8-10 (!7&!9)\|\|(!1&!9) | | |

Let us consider an example, in order to show how Table 2 should be interpreted. The line corresponding to tower T3 indicates that:
- T3 normally receives from link 3-4.
- If T1 is down, then T3 receives from link 2-3.
- If T4 and T6 are both down, or T4 and T10 are both down, then T3 receives from expensive link 3-5.
- T3 normally sends data through link 2-3.
- If T1 is down or T2 is down, then T3 sends data through link 3-4.
- If T2 and T4 are both down, or T2 and T10 are both down, then T3 sends through expensive link 1-3.
- If T1 and T4 are both down, then T3 sends through expensive link 3-5.

We formulated Table 2 and derived rules to be used in the formal model of the transmission line. As a result, we have listed the rules for normal, rerouted, and expensive send/receive communications in Table 3.

Table 3: Rules for normal, rerouted, and expensive communication.

| | Condition | Rule |
|---|---|---|
| **Send** | 1<i<10 | Normal send when all nodes are operational |
| | i=2 | Rerouted send when !$T_1$<br>Expensive send when !$T_1$ & !$T_3$ |
| | 2<i<6 | Rerouted send when !$T_1$ \|\| !$T_{i-1}$<br>Expensive send when (!$T_{i-1}$&!$T_{i+1}$)\|\|(!$T_{10}$&!$T_{i-1}$)<br>Expensive reverse send when (!$T_1$&!$T_{i+1}$) |
| | 5<i<9 | Rerouted send when !$T_{10}$ \|\| !$T_{i+1}$<br>Expensive send when (!$T_{i-1}$&!$T_{i+1}$)\|\|(!$T_1$&!$T_{i+1}$)<br>Expensive reverse send when (!$T_{10}$&!$T_{i-1}$) |
| | i=9 | Rerouted send when $!T_{10}$<br>Expensive send when $!T_8$ & $!T_{10}$ |
| Receive | 0<i<5 &<br>6<i<11 | Normal receive when all nodes are operational |
| | 0<i<3 | Expensive receive when $(!T_{i+1}\&!T_{i+3})||(!T_{10}\&!T_{i+1})$ |
| | i=3 | Rerouted receive when $!T_1$<br>Expensive receive when $(!T_{i+1}\&!T_{i+3})||(!T_{10}\&!T_{i+1})$ |
| | i=4 | Rerouted receive when $!T_1 ||  !T_{i-2}$<br>Expensive receive when $(!T_1\&!T_{i-1})||(!T_{10}\&!T_{i+1})$ |
| | 4<i<7 | Rerouted receive when $(!T_1\&!T_{i-2})||(!T_{10}\&!T_{i+2})$<br>Expensive receive when $(!T_1\&!T_{i-1})||(!T_{10}\&!T_{i+1})$ |
| | i=7 | Rerouted receive when $!T_{10} ||  !T_{i+2}$<br>Expensive receive when $(!T_1\&!T_{i-1})||(!T_{10}\&!T_{i+1})$ |
| | i=8 | Rerouted receive when $!T_{10}$<br>Expensive receive when $(!T_{i-3}\&!T_{i-1})||(!T_1\&!T_{i-1})$ |
| | 8<i<11 | Expensive receive when $(!T_{i-3}\&!T_{i-1})||(!T_1\&!T_{i-1})$ |

### 3.2. Formal Model Of The Transmission Line

*PRISM implementation of this model is listed in Appendix B.*

Actually, this is the formal model of a Transmission Line that has only 10 towers. Each tower in the line has a bone node. A bone node is called a terminal node if it is in either end of the line.

Formalization is a continuous-time Markov chain with rewards.

All bone nodes start being fully operational. A bone node may fail with a certain rate, and then get recovered with another certain rate.

### 3.2.1. Description of the bone node models

A **non-terminal bone node** maybe in four operational states:
State0: BROKEN:
- Waits to be recovered.
- State-Reward "$fail_i$" applies, for computing expected failure time of tower $T_i$.

State1: OPERATIONAL:
- A TX takes place.
- RX may take place.
- Device can break down.
- Transition-Rewards "$sentpacketsT_i$" and "$receivedpacketsT_i$" apply, for computing the expected number of sent and received packets from/at tower $T_i$.
- Transition-Reward "$backup_{ij}$" applies, for computing the expected number of usage of backup link $TX_{ij}$.

- Reward "battery$T_i$" applies, for computing the expected energy consumption caused by TX and RX transmissions.

State2: SLEEP:
- Device can wake up.
- Device can break down.
- Reward "battery$T_i$" applies, for the expected time spent in this state.

State3: DONE:
- RX may take place.
- Device can break down.
- Transition-Reward "receivedpackets$T_i$" applies, for computing the expected number of received packets at tower $T_i$.
- Reward "battery$T_i$" applies, for computing the expected energy consumption caused by RX transmissions.

The transition between these four states can be summarized as:
- duty cycle: SLEEP(2)->OPERATIONAL(1)->DONE(3)->SLEEP(2)
- break down: AnyMode(1,2,3) -> BROKEN(0)
- recovery:   BROKEN(0) -> SLEEP(2)

A ***terminal bone node*** may be in three states:

State0: BROKEN:
- Can be recovered.
- State-Reward "fail$_i$" applies, for computing expected failure time of tower $T_i$.

State1: OPERATIONAL:
- RX takes place.
- Device can break down.
- Transition-Reward "receivedpackets$T_i$" applies, for computing the expected number of received packets at tower $T_i$.
- Reward "battery$T_i$" applies, for computing the expected energy consumption caused by RX transmissions.

State2: SLEEP:
- Can wake up.
- Device can break down.
- Reward "battery$T_i$" applies, for the expected time spent in this state.

The transition between these four states can be summarized as:
- duty cycle: SLEEP(2)->OPERATIONAL(1)->LEEP(2)
- break down: AnyMode(1,2,3) -> BROKEN(0)
- recovery:   BROKEN(0) -> SLEEP(2)

### 3.2.2. Sanity check for the formal model

- All the devices start in sleep mode, and they synchronously:
    - wake up after a predefined sleeping time - *network becomes operational*
    - go to sleep after a predefined operation time – *network sleeps*

- When the network is in operational mode
  - All the non-terminal devices send data to their most convenient neighbor
  - Terminal devices always receive data, non-terminal devices only receive data if the routing should be done so
- If a device fails
  - while the network is in sleep mode, then the rest of the network should still wake up normally
  - while the network is in operational mode, then the rest of the network should still transmit data and sleep normally
- If a device is recovered
  - while the network is in sleep mode, then the recovered device will wake with the rest of the network
  - while the network is in operational mode, then the recovered device waits in sleep mode until the network goes to sleep mode and wakes up afterwards

### 3.2.3. Parameters

***P1.** Number of bone nodes (or towers)*
In this case study, we considered a network of 10 bone nodes. The number of bone nodes can be scaled up and down. However, this model already produces 590,848 reachable states. The calculation of the number of states is roughly the multiplication of $3^{tt}*4^{ntt}$ where *tt* is the number of terminal towers and *ntt* is the number of non-terminal towers. It is obvious that this model can not deal with the real case of 100 towers, since the number of states would then be around $10^{60}$.

***P2. rFail:** Failure rate per sensor*
It is possible to assign different failure rates to different sensors
***P3. rRecover:** Recovery rate per sensor*
It is possible to assign different recovery rates to different sensors
***P4. rSend:** Packet rate per sensor*
It is possible to assign different packet rates to different sensors
***P5. cSend:** Cost of a packet*
It is possible to assign different costs for different sensors

### 3.3. Analysis of the Model

In this section, we present a small set of possible answers we can learn from the analysis of the model. There are of course many more questions that can be asked to this model. See Section 4.2 for an exhaustive list.

#### 3.3.1 Failures

***M1. Expected number of sensor failures until time T***
R{"TotalNumberOfSensorsFailures"}=? [C<=T]

    0.9999001225814035    (T=100,000)
    9.999000327135414    (T=1,000,000)
    9.999000099990002E-6    (T=1)

***M2. Probability of a sensor failure within first T time units***
P=? [F<=T !s1|!s2|!s3|!s4|!s5|!s6|!s7|!s8|!s9|!s10]
    0.6321205588600892    (T=100,000)

### 3.4. Summary and Evaluation

In this model, we can reason about each tower's behaviour abstracting the intra-tower sensor network. We can measure the battery consumption in details. We can thoroughly analyze the behaviour of the routing algorithm., such as the usage details of the backup lines. Besides we can track failure and repair information, number of sent and received packets, and battery consumption. One drawback of the model is, battery replacement cannot be modelled. Another drawback is, since the model already includes too many details it is not possible to work on a realistic sized (e.g. 100 towers) network.

# 4. Electricity Tower

In this section, we only consider a single electricity tower. Therefore, we consider a single bone node with directly connected sensor nodes.

## 4.1. Formal Model Of A Single Tower

*PRISM implementation of this model is listed in Appendix C.*

Formalization is a continuous-time Markov chain with rewards. All sensors start being fully operational. A sensor may fail with a certain rate, and then get recovered with another certain rate.

### 4.1.1. Model Description

A sensor maybe in two operational states: operational and non-operational. In the operational mode, the sensor can send data and therefore consume energy.

Model needs to be as simple as possible in order to enable model checking. This calls for significant abstractions. For example:
- Energy consumption of a node cannot be reset after recovery. Similarly, a device does not stop operating after a certain threshold of energy.
- Energy consumption in sleep mode is omitted.
- There is no distinction between send and receive. The send action in the formal model corresponds to the combination of the send and receive operations in the real model (when the sensor wakes up). Therefore, the energy consumption in the send action is actually the sum of the consumptions in Tx and Rx.

### 4.1.2. Parameters

**P1.** *Number of sensors*
   Theoretically, we can handle maximum 36-37 sensors with model checking (assuming that PRISM MTBDD engine can handle models up to $10^{10}$ or $10^{11}$ states). As a simple example, a model of 10 sensors produces $2^{10}=1024$ states and $(2^{10} \times 11)-1=11263$ transitions, and almost all model checking operations finish under a second.

**P2. rFail:** *Failure rate per sensor*
   It is possible to assign different failure rates to different sensors

**P3. rRecover:** *Recovery rate per sensor*
   It is possible to assign different recovery rates to different sensors

**P4. rSend:** *Packet rate per sensor*
   It is possible to assign different packet rates to different sensors

**P5. cSend:** *Cost of a packet*
   It is possible to assign different costs for different sensors

## 4.2. Analysis of the Model

### 4.2.1. Failures

***M1. Expected number of sensor failures until time T***
R{"TotalNumberOfSensorsFailures"}=? [C<=T]
    0.9999001225814035    (T=100,000)
    9.999000327135414    (T=1,000,000)
    9.999000099990002E-6    (T=1)

***M2. Probability of a sensor failure within first T time units***
P=? [F<=T !s1|!s2|!s3|!s4|!s5|!s6|!s7|!s8|!s9|!s10]
    0.6321205588600892    (T=100,000)

***M3. Probability of no sensor fails (at all times) within T time units (1-M2)***
P=? [G<=T s1&s2&s3&s4&s5&s6&s7&s8&s9&s10]
    0.3678794411399108    (T=100000 )

***M4. Probability of a specific sensor failing at some point within first T time units***
P=? [F<=T !Si]
    0.009950166250892718 (T=10,000 Si=s1)

***M5. Expected number of recoveries until time T***
R{"TotalNumberOfRecoveries"}=? [C<=T]
    0.998001974867715    (T=100,000)

***M6. Long-run probability of two sensors being non-operational at the same time (Requires an additional module I called "failcount")***
S=? [failure=2]
    4.4955024745493276E-7
Note: Iterative methods Power and Jacobi did not converge within 10000 iterations, so we used Gauss-Seidel method.

***M7. Long-run probability of a sensor being non-operational (Requires an additional module I called "failcount")***
S=? [failure=1]
    9.990005498998502E-4

***M8. Probability of two sensors being non-operational at the same time within first T time units (Requires an additional module I called "failcount")***
P=? [F<=T failure=2]
    4.5079475214300257E-7    (T=120)
    8.086355995439531E-6    (T=1,000)
    8.892885869021347E-5    (T=10,000)

***M9. Probability of a sensor being non-operational within first T time units (Requires an additional module I called "failcount")***
P=? [F<=T failure=1]
       9.995001666255187E-4    (T=100)
       0.009950166250865224    (T=1,000)
       0.0951625819675857    (T=10,000)
       0.6321205589509482    (T=100,000)
       0.864664716914856    (T=200,000)

### 4.2.2. Energy

***M5. Expected number of packets that the bone node receives until time T***
R{"TotalNumberOfCommunicationsToBN"}=? [C<=T]
       999900.1225814035    (T=100,000)

***M6. Expected energy consumption of the bone node resulting from communication with the sensors until time T (M5xcost)***
CostOfCommunicationWithASensor x R{"TotalNumberOfCommunicationsToBN"}=? [C<=T]

**LESS MEANINGFUL QUESTIONS (in terms of results)**

***LM1. Expected number of INSTANT sensor failures in the long run (A special case of M1)***
R{"TotalNumberOfSensorsFailures"}=? [ S ]    9.999000099990002E-6

***LM2. Probability of a sensor failure within first T time units (A special case of M2)***
P=? [s1&s2&s3&s4&s5&s6&s7&s8&s9&s10 U<=T
!s1|!s2|!s3|!s4|!s5|!s6|!s7|!s8|!s9|!s10]    0.6321205588600892 (T=100,000)

***LM3. Expected cumulative energy consumption of a sensor at a certain point***
R{"s1"}=? [C<=T]

### 4.4. Summary and Evaluation

This is a simplistic model, designed to be able to cover more nodes. However, this simplicity is not enough for covering 50 sensor nodes in a tower. Though, for a limited number of nodes we can learn much information as seen in the previous section.

# 5. Compact Model

We present our stochastic model for the whole of the considered Chinese Smart Grid. We explain parts of our model, using source codes in PRISM description language placed in tables below.

The idea:
- We use the compact modelling approach that was developed in a PhD thesis[2].
- We keep track of a) *failed nodes*, and b) *operational states of nodes* in different variables.

## 5.1. Assumptions

### 5.1.1. Routing

We abstract the routing algorithm used in the network in order to suggest a more flexible and probabilistic model. We require the user to enter the probability of taking a cheap link (or an expensive link), instead of specifying which routing scheme is used. Therefore, we are able to verify our model for different probabilities and reason on the performance of the selected routing algorithm.

**Possible future work**
A future work arising from this abstraction is to model routing algorithms and produce realistic probability values.

### 5.1.2. Duty Cycle

In essence we assume that the bone nodes and the sensor nodes have different duty cycles. We think that these duty cycles resemble to the graphs in Figure 5 and Figure 6. A sensor node sleeps, then wakes up and starts sensing, then sends its data to the bone node (of its tree), then goes to sleep again. A bone node sleeps, then wakes up and starts receiving data from the sensor nodes (in its tree), then processes the data and routes[3] it to another tower (i.e. bone node), then goes to sleep again.

---

[2] E. Yuksel. Qualitative and Quantitative Security Analyses of ZigBee Wireless Sensor Networks

[3] Routing may have different meanings depending on the context. In this context, routing is an operation that includes two consecutive steps: 1) running a routing algorithm to determine which tower to forward data, and 2) send the data.
Notice that this is different than the usage in Section 2.1. which is merely the first step and thus excludes the "send" operation.

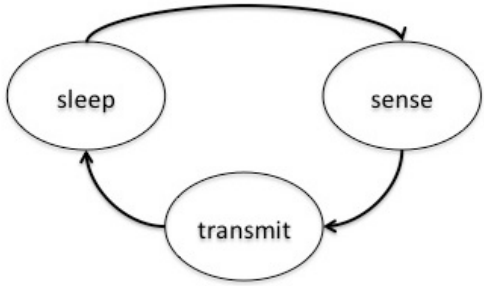
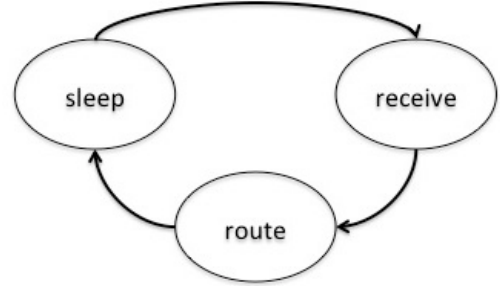

Figure 5: Duty cycle of a sensor node

Figure 6: Duty cycle of a bone node

As you can see, the duty cycles are very similar both in states and transitions. Our first abstraction is merging the no-sleep states (i.e. sense and transmit for sensor node, receive and route for bone node) and thus having just two states in the duty cycle: *sleep* and *process*. We call this as abstract duty cycle, as shown in Figure 7 together with action labels *wakeup* and *process* on top of transitions. Our second abstraction is, we assume that all nodes share the same duty cycle. In other words, all nodes are sleeping for the same period, and they are processing in the same period – they are *synchronized*.

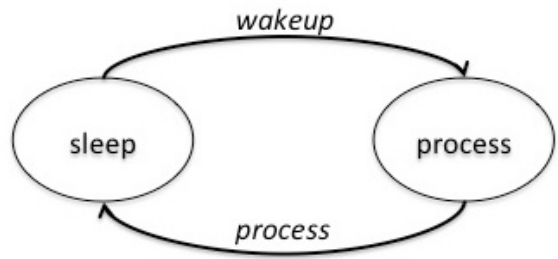

Figure 7: Abstract duty cycle for all nodes

An important point regarding duty cycle is *timing*. In fact, we don't know how much time is spent in each state, and we don't know if those times are constant. We prefer a stochastic model, such that we know mean sojourn time for each state but the delay is not deterministic. For example, if sensors sleep for 60 minutes in average, they may sleep for 57 minutes in the first cycle and maybe they will sleep for 61 minutes in the second. As explained in Section 1.1., the sleeping time is actually distributed with exponential distribution using the rate for sleep. The same principle applies for process state, as well. We believe that such a stochastic model is more convenient and more realistic then using deterministic delay, since the philosophy of wireless sensor networks can be described as "finish your job and go to sleep as soon as possible". Thus, each sensor should go to sleep as long as they are done, and the time required for that will not be the same for all sensors. Besides, clock drifts may happen which may lead to unwanted consequences.

### Possible future work

A future work arising from this point is improving the model by adding the possibility of a transition from state sense to the state sleep if the sensed data is not significant enough to transmit and therefore consume energy. This improvement however conflict with the current abstraction of having just two states, which are identical in both bone and sensor nodes.

### 5.1.3. Break-down and Recovery

We assume that any node can break down at some point, defined with a *fail* rate. We also assume that, after failure a team is sent to the transmission line to fix the broken nodes, defined with a *recover* rate. We assumed that a BN failure is much more critical than an SN failure, therefore the recovery team always fixes the BNs first. An important point in assumptions is, we set a threshold of failures such that if a certain number of nodes are failed at the same time, then the whole network is down. This threshold is distinct for BNs and SNs and can be adjusted according to new managerial decisions.

## 5.2. The Model

*PRISM implementation of this model is listed in Appendix D.*

### 5.2.1. Input Parameters

In Table 4, we have the constant definitions to be used in the model. These constants are classified in four groups: *number of nodes*, *durations*, *probabilities*, and *costs*. This is the only part in the model that we need data from the engineers/developers. The accuracy of the results heavily relies on the accuracy of the input parameters.

*Number of nodes* is obviously very influential in the overall size of the model. If we consider 100 towers (i.e. SIZE_BN=100) where each tower is equipped with 50 sensor nodes (i.e. SIZE_SN=50) and a single bone node, then in total we have 5100 nodes.

*Durations* are used in computing the rates to be used in the model. A duration is an "average" time for a certain activity, where time unit is an **hour**. We need duration values for a) duty cycle of the nodes (i.e. SLEEPTIME and PROCESSTIME), b) average uninterrupted lifetime of the nodes (i.e. MEANTIMEBETWEENFAILURE_SN and MEANTIMEBETWEENFAILURE_BN), and c) average recovery time in case of failure (i.e. RECOVERYTIME_SN and RECOVERYTIME_BN).

Currently, the mere *probability* used in the model (i.e. pCHEAPLINK) is the probability of using a regular (cheap) link. In other words, this is the point where we are asking the user to specify the routing scheme (*see Section 2.1.*).

Costs are the energy consumption values for certain situations. Currently, we employ costs for using cheap and expensive links, intra-tower communication, processing and sleep modes.

**Table 4: Input parameters**

```
// 1. Number of nodes
const int SIZE_BN=100;                              // Number of bone nodes
const int MAX_BN_FAIL=5;                            // MAX number of bone node failures
const int SIZE_SN=50;                               // Number of sensor nodes in each tower
const int MAX_SN_FAIL=50;                           // MAX number of sensor node failures that can be tolerated
// 2. Durations
const int SLEEPTIME=1;                              // Average sleeping duration for each node (for example: 1 hour)
const int MEANTIMEBETWEENFAILURE_SN=24000;          // Average lifetime without failure per sensor node (for example: 1000 days)
const int MEANTIMEBETWEENFAILURE_BN=36000;          // Average lifetime without failure per bone node (for example: 1500 days)
const int RECOVERYTIME_SN=48;                       // Average recovery time for sensor nodes(for example: 48 hours)
const int RECOVERYTIME_BN=36;                       // Average recovery time for bone nodes (for example: 36 hours)
const double PROCESSTIME=0.001;                     // Average processing duration (for example: 3.6 seconds)
// 3. Probabilities
const double pCHEAPLINK=0.95;                       // Probability of using the inexpensive transmission links
// 4. Costs
const int cCHEAPTX=24;                              // Cost of inter-tower + intra-tower comm. when using an
                                                    // inexpensive tranmsission link
const int cEXPENSIVETX=40;                          // Cost of inter-tower + intra-tower comm. when using an expensive
                                                    // expensive tranmsission link
const int cSNTX=8;                                  // Cost of intra-tower comm. for a sensor node
const double cSLEEP_BN=0.001;                       // Cost of sleep for a bone node
const double cSLEEP_SN=0.001;                       // Cost of sleep for a sensor node
const int cPROCESS_BN=5;                            // Cost of processing for a bone node
const int cPROCESS_SN=2;                            // Cost of processing for a sensor node
```

### 5.2.2. Rates

As we are modelling a CTMC, we need to produce rates (see Section 1.1.) from the input parameters of the model. This is simply done by inverting all the durations defined in Table 4. The resulting rates are presented in Table 5.

**Table 5: Rates**

```
const double rSLEEP=1/SLEEPTIME;
const double rFAIL_SN=1/MEANTIMEBETWEENFAILURE_SN;
const double rFAIL_BN=1/MEANTIMEBETWEENFAILURE_BN;
const double rRECOVERY_SN=1/RECOVERYTIME_SN;
const double rRECOVERY_BN=1/RECOVERYTIME_BN;
const double rPROCESS=1/PROCESSTIME;

formula osnf = 1-(0.01*(failedSN/(SIZE_SN*SIZE_BN)));
formula obnf = 1-(0.01*(failedBN/SIZE_BN));
```

We have designed the rates to be dynamic depending on the size of the active devices on the network. Therefore, we developed formulae that we named operational node factor (i.e. osnf, obnf).

### 5.2.3. Controller Module

The controller module is an auxiliary module in the model that implements synchronization between the nodes. The controller module is shown in Table 6, where we can see a local variable (named mode), followed by two commands (labeled with actions wakeup and process, respectively).

When the nodes are asleep, mode variable has the value of 1; and when they are awake (or processing), mode has value of 2.

As mentioned before, PRISM's guarded commands have a simple form comprising a guard and updates. The first command, wakeup, defines the guard that should be satisfied for transition from *sleep* state to *process* state. Obviously, the nodes should be in the sleep state in order to wake up. But we have a stronger guard, such that if maximum tolerated number of sensor nodes fail, or maximum tolerated number of bone nodes fail, then the guard is not satisfied. The guard for processing is also similarly formed.

Updates of the two commands in the controller module take place with the rates rSLEEP and rPROCESS, respectively.

**Table 6: Controller Module**

```
module controller
        mode : [1..2] init 1;      // Modes for nodes:
                                   // 1: Sleep
                                   // 2: Process
        // WAKEUP
        [wakeup] mode=1 & failedBN<MAX_BN_FAIL & failedSN<MAX_SN_FAIL -> rSLEEP: (mode'=2);

        // SENSE-AND-SEND OR RECEIVE-AND-ROUTE
        [process] mode=2 & failedBN<MAX_BN_FAIL & failedSN<MAX_SN_FAIL-> rPROCESS: (mode'=1);
endmodule
```

### 5.2.4. Sensor Nodes Module

Sensor Nodes module captures the breakdown and recovery of all the sensor nodes. It is the implementation of breakdown and recovery model defined in Section 2.3. The corresponding Prism model is given in Table 7.

The guards in this module are asserting that, a) in order a failure to take place, there should be at least one operational node, b) in order a recovery to take place, there should be at least one failure in the network.

**Table 7: Sensor Nodes Module**

```
module sensorNodes

        failedSN : [0..MAX_SN_FAIL] init 0; // Number of failed sensor in this group

        // FAILURE
        [] failedSN<MAX_SN_FAIL-> osnf*rFAIL_SN: (failedSN'=failedSN+1);
        // RECOVERY
        [] failedSN>0 -> rRECOVERY_SN: (failedSN'=0);

endmodule
```

### 5.2.5. Bone Nodes Module

Bone Nodes module behaves similar to the Sensor Nodes module, except that it keeps track of bone nodes. As shown in Table 8, the maximum number of bone nodes, the rates for breakdown and recovery, and the local variable that keeps track of failed bone nodes are creating the difference.

**Table 8: Bone Nodes Module**

```
module boneNodes

        failedBN : [0..MAX_BN_FAIL] init 0; // Number of failed sensor in this group

        // FAILURE
        [] failedBN<MAX_BN_FAIL-> obnf*rFAIL_BN: (failedBN'=failedBN+1);
        // RECOVERY
        [] failedBN>0 -> rRECOVERY_BN: (failedBN'=0);

endmodule
```

## 5.4. Analysis of the Model

We can specify various properties using our compact model. Below we present a set of property specifications, and sample results for our case study. Besides, we present graphical results this time in order to see the trends depending on the changes in parameters.

### 5.4.1. Probability of Node Failure in the Long Run

We start our analyses with an example of how we can benefit from formal verification prior to deployment. Let us assume that we can get an estimation of Mean Time Between Failure (MTBF) for the sensor devices e.g. from vendors. Using this information and our formal model with the property **S=? [failedSN>0 | failedBN>0]**, we can compute the probability of any node failure in the long run.

We present the results for a range of MTBF values in Figure 8, in order to demonstrate how this value affects the node failure in the long run. We considered MTBF values between 250 and 1500 days, and analyzed the cases for both types of nodes. An interesting conclusion from this analysis is the influence of MTBF is diminished for larger MTBF values.

In addition, we are able to distinguish between node types and compute failure risks per node type. In this case study, the failure risk of any bone nodes is computed as 0.00099 whereas the same risk is 0.002 for any sensors in the system.

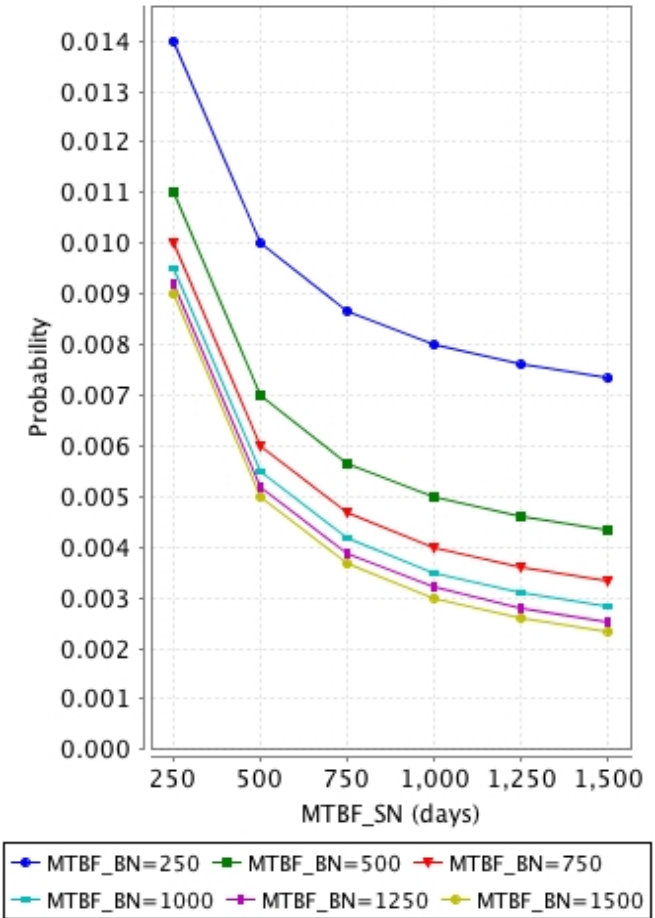
**Figure 8: Probability of node failure in the long run**

We present the numerical results in addition to the graphical results below.

**MTBF_BN=250**
| Time | Probability |
|---|---|
| 250.0 | 0.013999996247822541 |
| 500.0 | 0.00999999634575818 |
| 750.0 | 0.00866666303048286 |
| 1000.0 | 0.007999996370126678 |
| 1250.0 | 0.007599996373045063 |
| 1500.0 | 0.007333329707962957 |

**MTBF_BN=500**
| Time | Probability |
|---|---|
| 250.0 | 0.010999998967576006 |
| 500.0 | 0.006999999064993425 |
| 750.0 | 0.005666665749622834 |
| 1000.0 | 0.004999999089233332 |
| 1250.0 | 0.004599999092136408 |
| 1500.0 | 0.004333332427045969 |

**MTBF_BN=750**
Time          Probability

| | |
|---|---|
| 250.0 | 0.00999999946971127 |
| 500.0 | 0.005999999566956941 |
| 750.0 | 0.0046666662515547986 |
| 1000.0 | 0.003999999591154281 |
| 1250.0 | 0.003599999594052247 |
| 1500.0 | 0.0033333329289590105 |

**MTBF_BN=1000**

| Time | Probability |
|---|---|
| 250.0 | 0.009499999645288928 |
| 500.0 | 0.005499999742448836 |
| 750.0 | 0.004166666427030887 |
| 1000.0 | 0.003499999766624894 |
| 1250.0 | 0.003099999769520343 |
| 1500.0 | 0.002833333104425712 |

**MTBF_BN=1250**

| Time | Probability |
|---|---|
| 250.0 | 0.009199999726522063 |
| 500.0 | 0.005199999823630798 |
| 750.0 | 0.0038666650820337 |
| 1000.0 | 0.003199999847794116 |
| 1250.0 | 0.0027999998506880288 |
| 1500.0 | 0.0025333331855925435 |

**MTBF_BN=1500**

| Time | Probability |
|---|---|
| 250.0 | 0.008999999770640447 |
| 500.0 | 0.00499999986771498 |
| 750.0 | 0.003666666552281334 |
| 1000.0 | 0.0029999998918698125 |
| 1250.0 | 0.0025999998947627122 |
| 1500.0 | 0.002333333229666673 |

### 5.4.2. Impact of Repair Service on the Failure Risk

After the deployment, we are always able to improve (or sometimes impair) capabilities such as repairing failed nodes. Assuming that we have deployed sensors and bone nodes having MTBFs of 1000 and 1500 days, respectively (which results in a failure probability of slightly less than 0.003), and our average recovery times were 48 hours for a sensor node and 36 hours for a bone node, we can compute the effects of our repair service performance for a range of repair times. We use the same property specification as before, however this time recovery times are variable.

In Figure 9, we present the results for average recovery times (abbreviated as RT) ranging from 12 to 72 hours. We can interpret these results as the recovery times have a significant effect on the robustness of the system, reminding the policy that bone nodes are more important and therefore have a high priority in repair service. However, the relation between the performance of the recovery service and the overall failure probability is more balanced, in the sense that the more improved service means the less probability of failure, and vice versa.

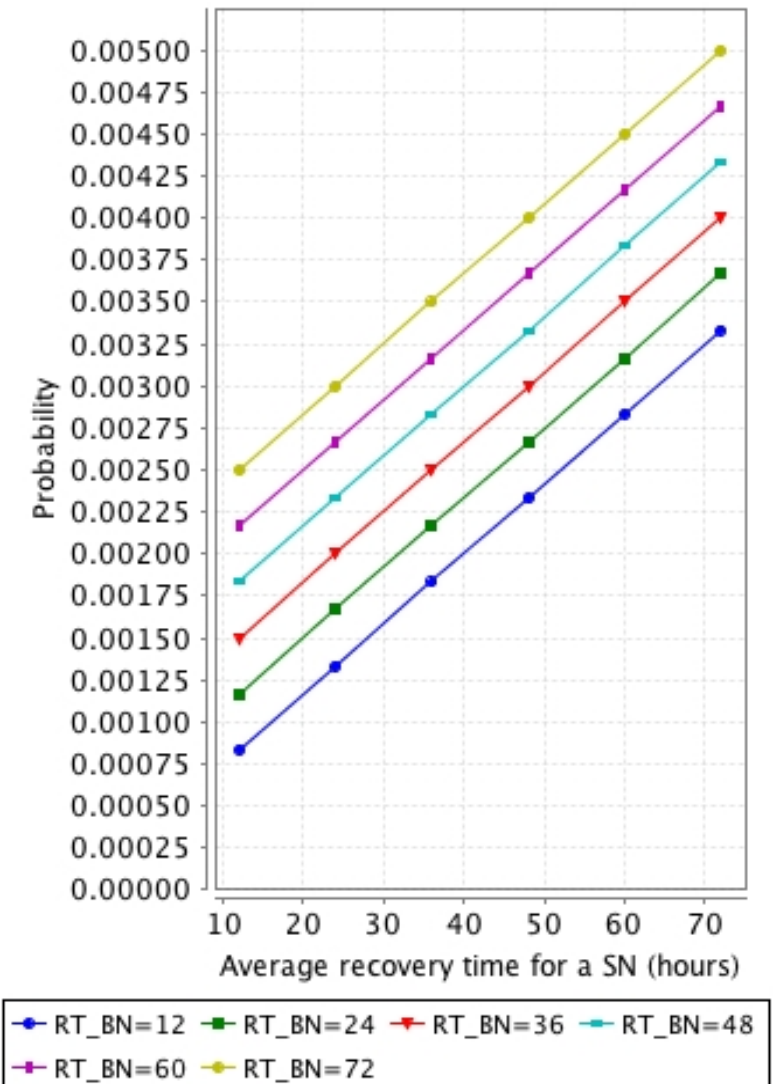

**Figure 9: Risk of failure depending on recovery times**

Again, we present the numerical results below.

**RT_BN= 12 hours**
RT_SN Probability
12.0   8.333333217179349E-4
24.0   0.0013333333202155153
36.0   0.0018333333177097684
48.0   0.0023333333141991653
60.0   0.0028333333096821674
72.0   0.0033333333041573217

**RT_BN= 24 hours**
RT_SN Probability

| 12.0 | 0.0011666666216913243 |
| 24.0 | 0.0016666666201878974 |
| 36.0 | 0.0021666666176807944 |
| 48.0 | 0.0026666666141685357 |
| 60.0 | 0.0031666666609649534 |
| 72.0 | 0.0036666666604122396 |

**RT_BN= 36 hours**
RT_SN Probability
| 12.0 | 0.001499999899397662 |
| 24.0 | 0.0019999998978929077 |
| 36.0 | 0.002499999895384126 |
| 48.0 | 0.0029999998918698125 |
| 60.0 | 0.0034999998873484696 |
| 72.0 | 0.003999999881818639 |

**RT_BN= 48 hours**
RT_SN Probability
| 12.0 | 0.001833333154814647 |
| 24.0 | 0.002333333153308199 |
| 36.0 | 0.0028333331507974266 |
| 48.0 | 0.003333333147280786 |
| 60.0 | 0.0038333331427567864 |
| 72.0 | 0.004333333137223828 |

**RT_BN= 60 hours**
RT_SN Probability
| 12.0 | 0.0021666663879199386 |
| 24.0 | 0.0026666663864114977 |
| 36.0 | 0.0031666663838983643 |
| 48.0 | 0.0036666663803790554 |
| 60.0 | 0.004166666375852004 |
| 72.0 | 0.0046666663703156435 |

**RT_BN= 72 hours**
RT_SN Probability
| 12.0 | 0.0024999995986911875 |
| 24.0 | 0.002999999597180384 |
| 36.0 | 0.003499999594664544 |
| 48.0 | 0.0039999995911422035 |
| 60.0 | 0.004499999586611872 |
| 72.0 | 0.004999999581071811 |

### 5.4.3. Expected Energy Consumption

Another question is what is the best time for replacing the batteries of the nodes. In order to estimate a property time for replacement, we verified the power consumption of devices using the properties **R{AvgEnergyBN}=? [C<=24*T1]** and **R{AvgEnergySN}=? [C<=24*T1]**. According to our abstract energy consumption model, we compute the expected consumption values for a bone node and a sensor node in average. Assuming that all nodes use batteries with the same capacity (e.g. 200,000 units), we can assist the system management in determining a convenient time for replacement which should take place before the batteries are totally drained.

We present an example in Figure 10 where we propose replacing bone node batteries every 300 days and sensor batteries every 900 days according to verification results.

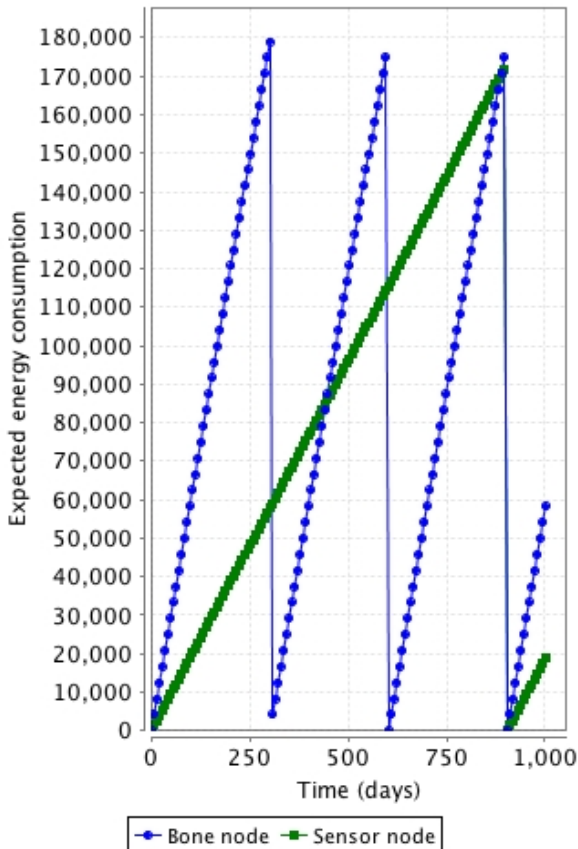

**Figure 10: Determining battery replacement times**

The numerical results of this analysis is presented below:

## Sensor node

| Time | Exp. consumption | Time | Exp. consumption | Time | Exp. consumption |
|---|---|---|---|---|---|
| 0.0 | 0.0 | 336.0 | 64471.72028 | 672.0 | 128943.4411 |
| 7.0 | 1343.160845 | 343.0 | 65814.88112 | 679.0 | 130286.602 |
| 14.0 | 2686.321684 | 350.0 | 67158.04196 | 686.0 | 131629.7628 |
| 21.0 | 4029.482523 | 357.0 | 68501.20280 | 693.0 | 132972.9237 |
| 28.0 | 5372.643362 | 364.0 | 69844.36364 | 700.0 | 134316.0845 |
| 35.0 | 6715.804201 | 371.0 | 71187.52479 | 707.0 | 135659.2454 |
| 42.0 | 8058.965041 | 378.0 | 72530.68564 | 714.0 | 137002.4062 |
| 49.0 | 9402.125880 | 385.0 | 73873.84648 | 721.0 | 138345.567 |
| 56.0 | 10745.28671 | 392.0 | 75217.00733 | 728.0 | 139688.7279 |
| 63.0 | 12088.44755 | 399.0 | 76560.16817 | 735.0 | 141031.8887 |
| 70.0 | 13431.60839 | 406.0 | 77903.32902 | 742.0 | 142375.0496 |
| 77.0 | 14774.76923 | 413.0 | 79246.48986 | 749.0 | 143718.2104 |
| 84.0 | 16117.93007 | 420.0 | 80589.65071 | 756.0 | 145061.3713 |
| 91.0 | 17461.09091 | 427.0 | 81932.81155 | 763.0 | 146404.5321 |
| 98.0 | 18804.25175 | 434.0 | 83275.9724 | 770.0 | 147747.693 |
| 105.0 | 20147.41259 | 441.0 | 84619.13324 | 777.0 | 149090.8538 |
| 112.0 | 21490.57343 | 448.0 | 85962.29409 | 784.0 | 150434.0146 |
| 119.0 | 22833.73427 | 455.0 | 87305.45493 | 791.0 | 151777.1755 |
| 126.0 | 24176.89511 | 462.0 | 88648.61578 | 798.0 | 153120.3363 |
| 133.0 | 25520.05595 | 469.0 | 89991.77662 | 805.0 | 154463.4972 |
| 140.0 | 26863.21678 | 476.0 | 91334.93747 | 812.0 | 155806.658 |
| 147.0 | 28206.37762 | 483.0 | 92678.09831 | 819.0 | 157149.8189 |
| 154.0 | 29549.53846 | 490.0 | 94021.25916 | 826.0 | 158492.9797 |
| 161.0 | 30892.69930 | 497.0 | 95364.42 | 833.0 | 159836.1406 |
| 168.0 | 32235.86014 | 504.0 | 96707.58085 | 840.0 | 161179.3014 |
| 175.0 | 33579.02098 | 511.0 | 98050.74169 | 847.0 | 162522.4623 |
| 182.0 | 34922.18182 | 518.0 | 99393.90254 | 854.0 | 163865.6231 |
| 189.0 | 36265.34266 | 525.0 | 100737.0634 | 861.0 | 165208.7839 |
| 196.0 | 37608.50350 | 532.0 | 102080.2242 | 868.0 | 166551.9448 |
| 203.0 | 38951.66434 | 539.0 | 103423.3851 | 875.0 | 167895.1056 |
| 210.0 | 40294.82518 | 546.0 | 104766.5459 | 882.0 | 169238.2665 |
| 217.0 | 41637.98602 | 553.0 | 106109.7068 | 889.0 | 170581.4273 |
| 224.0 | 42981.14685 | 560.0 | 107452.8676 | 896.0 | 171924.5882 |
| 231.0 | 44324.30769 | 567.0 | 108796.0285 | 903.0 | 0.0 |
| 238.0 | 45667.46853 | 574.0 | 110139.1893 | 910.0 | 1343.160845 |
| 245.0 | 47010.62937 | 581.0 | 111482.3501 | 917.0 | 2686.32169 |
| 252.0 | 48353.79021 | 588.0 | 112825.511 | 924.0 | 4029.482535 |
| 259.0 | 49696.95105 | 595.0 | 114168.6718 | 931.0 | 5372.64338 |
| 266.0 | 51040.11189 | 602.0 | 115511.8327 | 938.0 | 6715.804225 |
| 273.0 | 52383.27273 | 609.0 | 116854.9935 | 945.0 | 8058.96507 |
| 280.0 | 53726.43357 | 616.0 | 118198.1544 | 952.0 | 9402.125915 |
| 287.0 | 55069.59441 | 623.0 | 119541.3152 | 959.0 | 10745.28676 |
| 294.0 | 56412.75525 | 630.0 | 120884.4761 | 966.0 | 12088.44761 |
| 301.0 | 57755.91608 | 637.0 | 122227.6369 | 973.0 | 13431.60845 |
| 308.0 | 59099.07692 | 644.0 | 123570.7977 | 980.0 | 14774.7693 |
| 315.0 | 60442.23776 | 651.0 | 124913.9586 | 987.0 | 16117.93014 |
| 322.0 | 61785.39860 | 658.0 | 126257.1194 | 994.0 | 17461.09099 |
| 329.0 | 63128.55944 | 665.0 | 127600.2803 | 1001.0 | 18804.25183 |

**Bone node**

| Time | Exp. consumption | Time | Exp. consumption | Time | Exp. consumption |
|------|------------------|------|------------------|------|------------------|
| 0.0 | 0.0 | 343.0 | 24979.46854 | 686.0 | 49958.93707 |
| 7.0 | 4163.244775 | 350.0 | 29142.71329 | 693.0 | 54122.18183 |
| 14.0 | 8326.489530 | 357.0 | 33305.95805 | 700.0 | 58285.42658 |
| 21.0 | 12489.73428 | 364.0 | 37469.2028 | 707.0 | 62448.67134 |
| 28.0 | 16652.97904 | 371.0 | 41632.44756 | 714.0 | 66611.9161 |
| 35.0 | 20816.22379 | 378.0 | 45795.69232 | 721.0 | 70775.16085 |
| 42.0 | 24979.46855 | 385.0 | 49958.93707 | 728.0 | 74938.40561 |
| 49.0 | 29142.71330 | 392.0 | 54122.18183 | 735.0 | 79101.65036 |
| 56.0 | 33305.95806 | 399.0 | 58285.42658 | 742.0 | 83264.89512 |
| 63.0 | 37469.20281 | 406.0 | 62448.67134 | 749.0 | 87428.13988 |
| 70.0 | 41632.44757 | 413.0 | 66611.9161 | 756.0 | 91591.38463 |
| 77.0 | 45795.69232 | 420.0 | 70775.16085 | 763.0 | 95754.62939 |
| 84.0 | 49958.93708 | 427.0 | 74938.40561 | 770.0 | 99917.87414 |
| 91.0 | 54122.18183 | 434.0 | 79101.65036 | 777.0 | 104081.1189 |
| 98.0 | 58285.42659 | 441.0 | 83264.89512 | 784.0 | 108244.3637 |
| 105.0 | 62448.67134 | 448.0 | 87428.13988 | 791.0 | 112407.6084 |
| 112.0 | 66611.91610 | 455.0 | 91591.38463 | 798.0 | 116570.8532 |
| 119.0 | 70775.16085 | 462.0 | 95754.62939 | 805.0 | 120734.0979 |
| 126.0 | 74938.40561 | 469.0 | 99917.87414 | 812.0 | 124897.3427 |
| 133.0 | 79101.65036 | 476.0 | 104081.1189 | 819.0 | 129060.5874 |
| 140.0 | 83264.89512 | 483.0 | 108244.3637 | 826.0 | 133223.8322 |
| 147.0 | 87428.13987 | 490.0 | 112407.6084 | 833.0 | 137387.0769 |
| 154.0 | 91591.38463 | 497.0 | 116570.8532 | 840.0 | 141550.3217 |
| 161.0 | 95754.62939 | 504.0 | 120734.0979 | 847.0 | 145713.5665 |
| 168.0 | 99917.87414 | 511.0 | 124897.3427 | 854.0 | 149876.8112 |
| 175.0 | 104081.1189 | 518.0 | 129060.5874 | 861.0 | 154040.056 |
| 182.0 | 108244.3636 | 525.0 | 133223.8322 | 868.0 | 158203.3007 |
| 189.0 | 112407.6084 | 532.0 | 137387.0769 | 875.0 | 162366.5455 |
| 196.0 | 116570.8531 | 539.0 | 141550.3217 | 882.0 | 166529.7902 |
| 203.0 | 120734.0979 | 546.0 | 145713.5665 | 889.0 | 170693.035 |
| 210.0 | 124897.3426 | 553.0 | 149876.8112 | 896.0 | 174856.2798 |
| 217.0 | 129060.5874 | 560.0 | 154040.056 | 903.0 | 0.0 |
| 224.0 | 133223.8321 | 567.0 | 158203.3007 | 910.0 | 4163.244756 |
| 231.0 | 137387.0769 | 574.0 | 162366.5455 | 917.0 | 8326.489512 |
| 238.0 | 141550.3216 | 581.0 | 166529.7902 | 924.0 | 12489.73427 |
| 245.0 | 145713.5664 | 588.0 | 170693.035 | 931.0 | 16652.97902 |
| 252.0 | 149876.8112 | 595.0 | 174856.2798 | 938.0 | 20816.22378 |
| 259.0 | 154040.0559 | 602.0 | 0.0 | 945.0 | 24979.46854 |
| 266.0 | 158203.3007 | 609.0 | 4163.244756 | 952.0 | 29142.71329 |
| 273.0 | 162366.5454 | 616.0 | 8326.489512 | 959.0 | 33305.95805 |
| 280.0 | 166529.7902 | 623.0 | 12489.73427 | 966.0 | 37469.2028 |
| 287.0 | 170693.0349 | 630.0 | 16652.97902 | 973.0 | 41632.44756 |
| 294.0 | 174856.2797 | 637.0 | 20816.22378 | 980.0 | 45795.69232 |
| 301.0 | 179019.5244 | 644.0 | 24979.46854 | 987.0 | 49958.93707 |
| 308.0 | 4163.244756 | 651.0 | 29142.71329 | 994.0 | 54122.18183 |
| 315.0 | 8326.489512 | 658.0 | 33305.95805 | 1001.0 | 58285.42658 |
| 322.0 | 12489.73427 | 665.0 | 37469.2028 | | |
| 329.0 | 16652.97902 | 672.0 | 41632.44756 | | |
| 336.0 | 20816.22378 | 679.0 | 45795.69232 | | |

# APPENDIX

## A. A Routing Scheme

In this section, we present a routing scheme that may be used in a transmission line.

The idea is to prefer *processing* to *communication*. In other words, the usage of expensive communication – namely the backup links – should be *the last resort*. Instead, more hops therefore more processing of data is suggested.

We would like to demonstrate this scheme on a sample transmission line of 10 bone nodes as shown in Figure 11. In this example, T1 and T10 are terminal bone nodes, which merely receive data from other bone nodes. A non-terminal bone node may route its data to either of the terminal bone nodes. The solid lines in the figure represent the regular (cheap) links, whereas the dashed lines represent the backup (expensive) links.

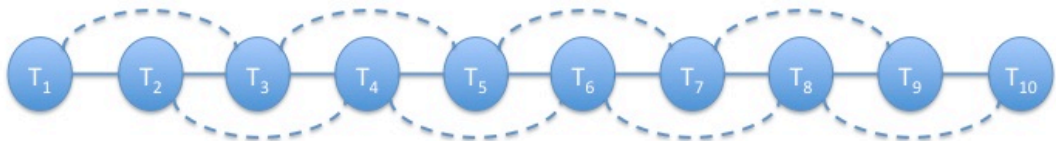

**Figure 11: Example topology, composed of 10 bone nodes.**

When all bone nodes are operational, the situation will be as shown in Figure 12. In this figure, and all the remaining figures in this section we will use the same format: 8 copies of transmission line, top-down arranged, each of them showing the behavior of the bone node colored in green. The reason of having 8 copies is we are only considering the non-terminal nodes. For example, in Figure 12 the top line is considering the bone node 2, and its route to a terminal bone node is also colored in green solid line. The second line is considering the bone node 3, and proceeding in this manner, the bottom line is considering the bone node 9. Obviously, the closest terminal bone node is chosen for routing when all bone nodes are operational. Therefore bone nodes 2, 3, 4, and 5 route to terminal bone node 1; whereas bone nodes 6, 7, 8, and 9 route to terminal bone node 10.

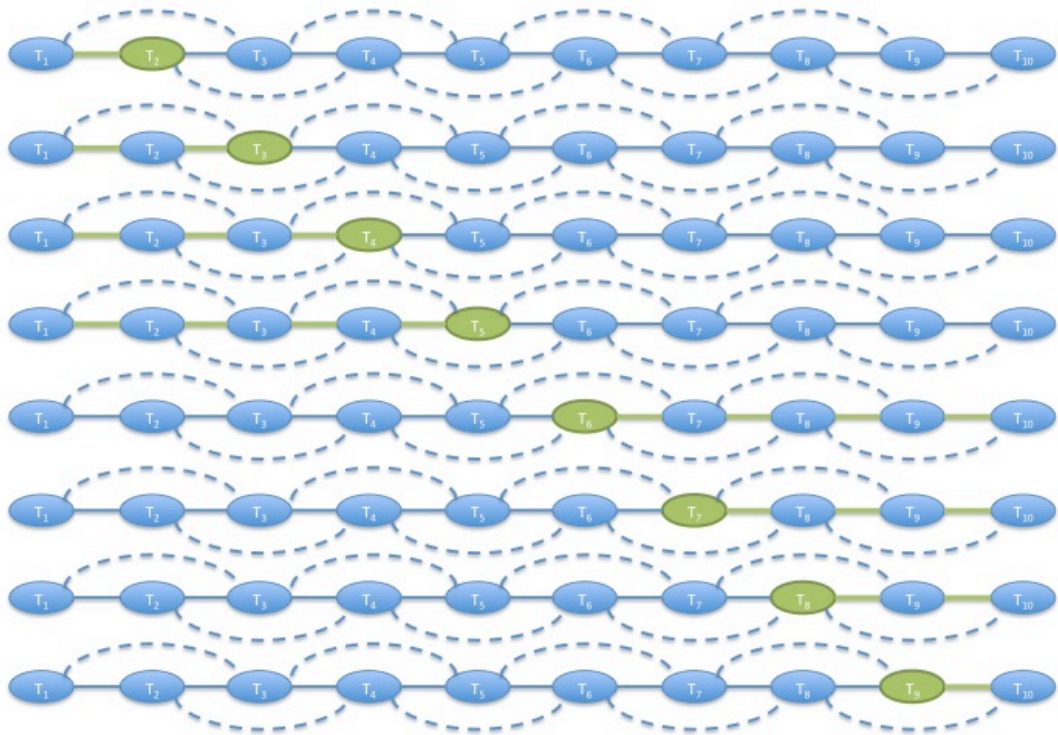

**Figure 12: All bone nodes are operational.**

Let us consider a case where a terminal bone node fails. In Figure 13, we demonstrate the case when bone node 1 fails. In this example, there is only one terminal bone node to receive data. Therefore, all the bone nodes route their data to bone node 10.

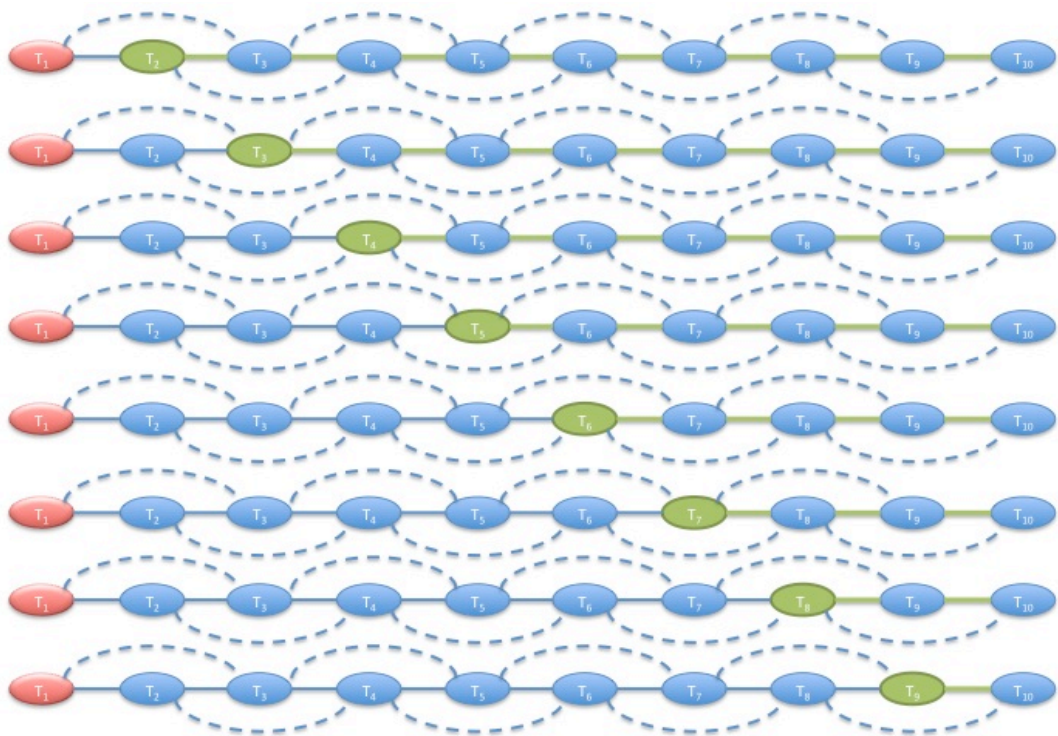

**Figure 13: Bone node 1 has failed.**

We can continue with considering a case where a single non-terminal bone node fails. In Figure 14, we present such a case where bone node 3 fails. This example allows us to demonstrate the idea of this routing scheme better. In the example. the bone node 4 reroutes its data to reverse direction and targets bone node 10, instead of using the expensive backup link to the bone node 2. Of course, in this case the bone nodes 5 to 10 receive extra data and do extra processing, but since processing is less costly than communication, it is perfectly fine. Obviously, rerouting also applies to bone node 5, which now routes towards the terminal bone node 10 just like bone node 4.

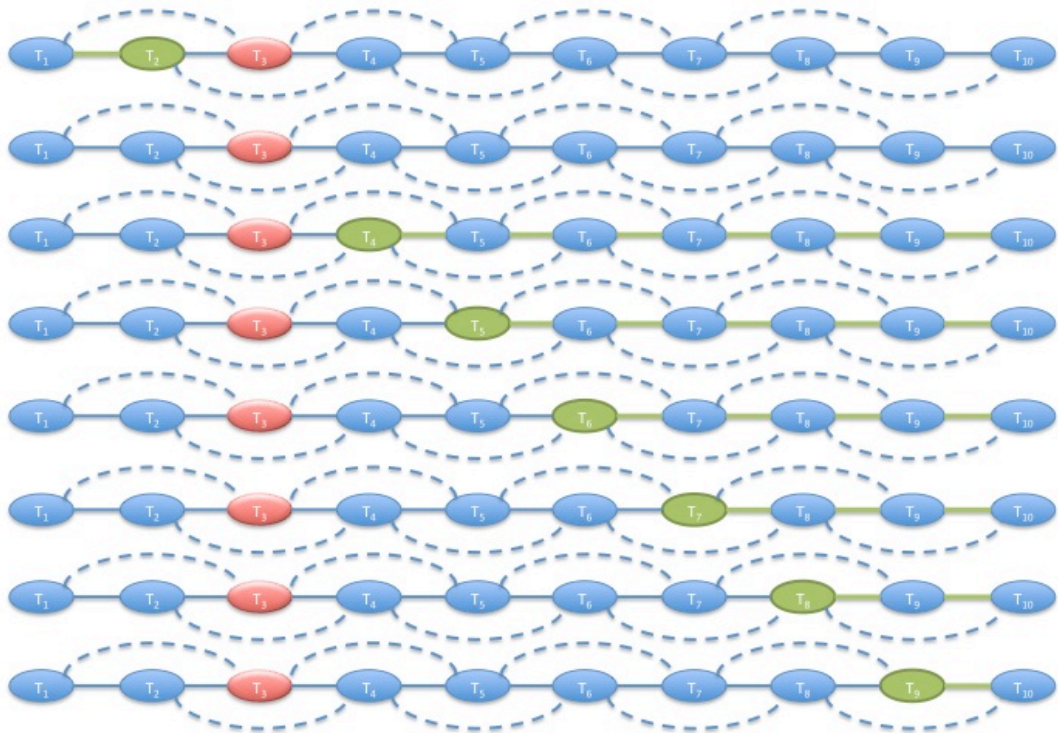

**Figure 14: Bone node 3 has failed.**

Let us consider a case where the usage of expensive backup links is impossible to avoid. An example could be when two neighbours of a bone node fail simultaneously as demonstrated in Figure 15. In this example, the bone node 2 has to use a backup link to the bone 4 as shown in green colored dashed line, since both bone node 1 and bone node 3 have failed.

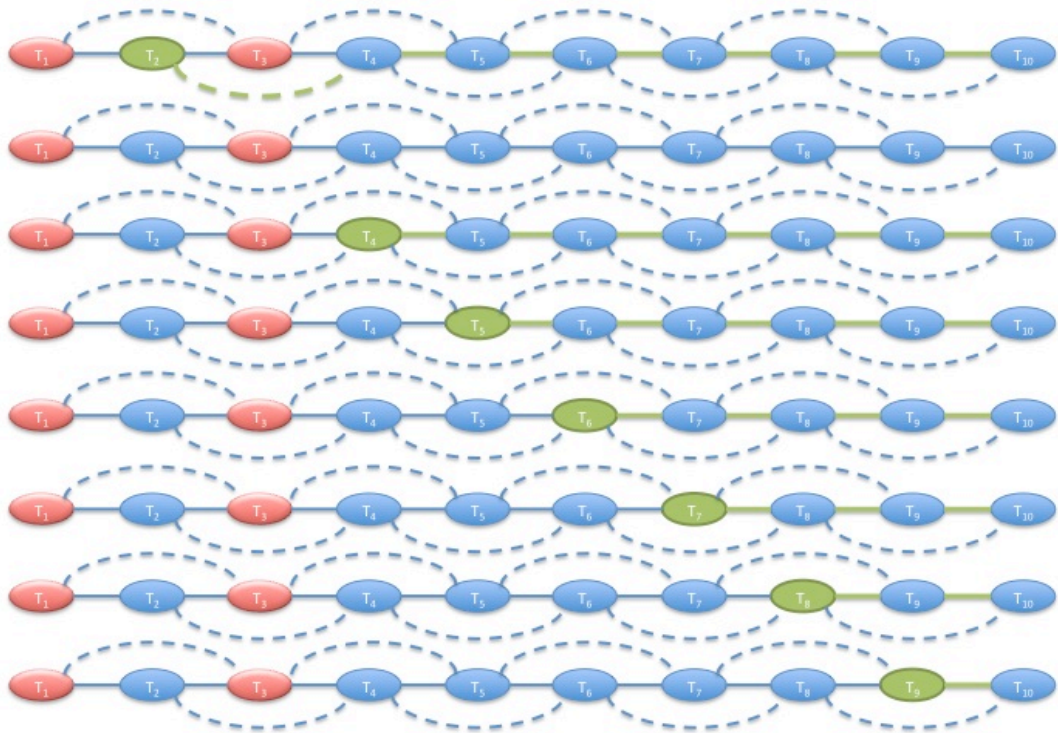

**Figure 15: Bone nodes 1 and 3 have failed.**

In the remaining figures below, we consider different cases in order to demonstrate how the routing strategy behaves in various failure conditions.

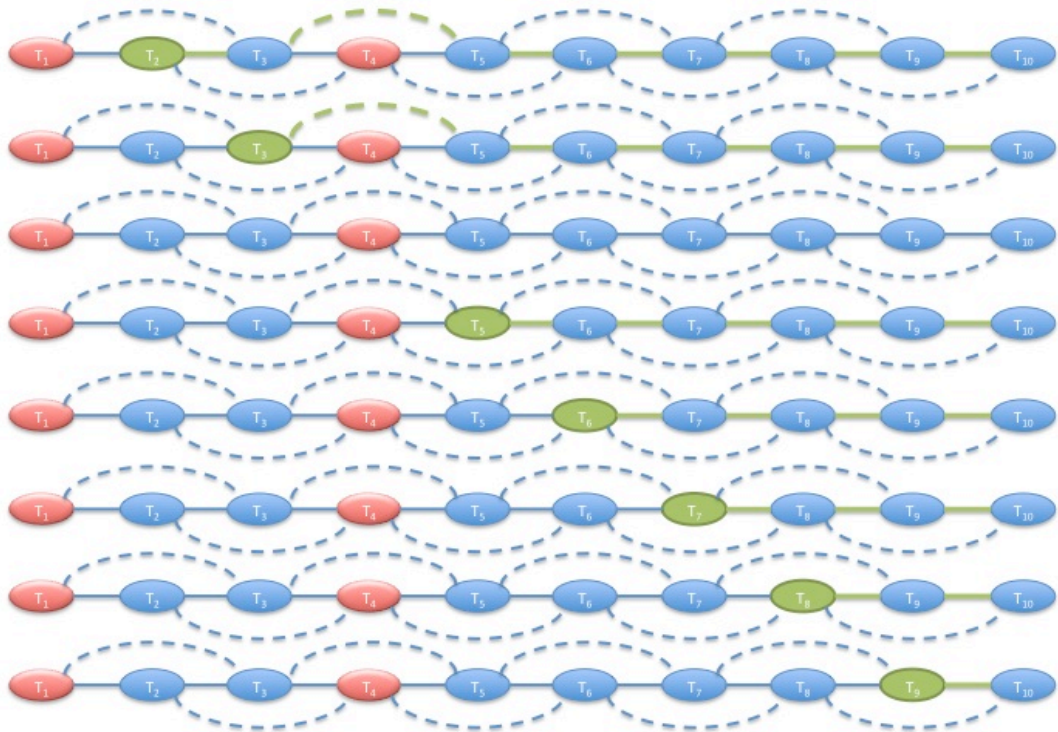

**Figure 16: Bone nodes 1 and 4 have failed.**

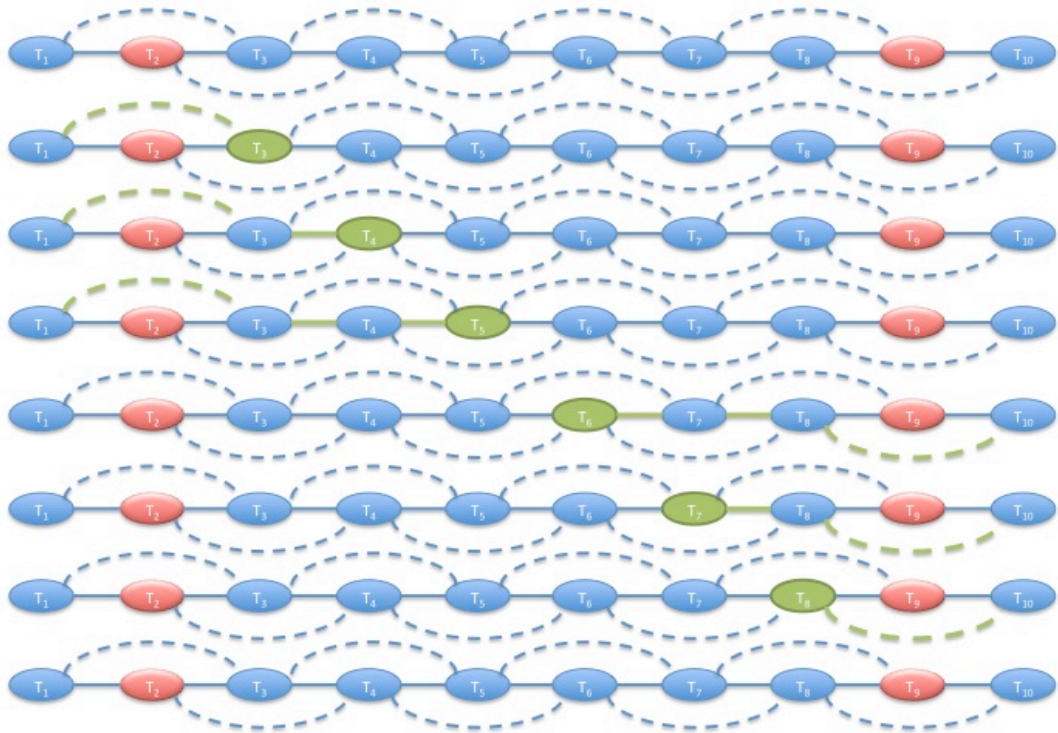

**Figure 17: Bone nodes 2 and 9 have failed.**

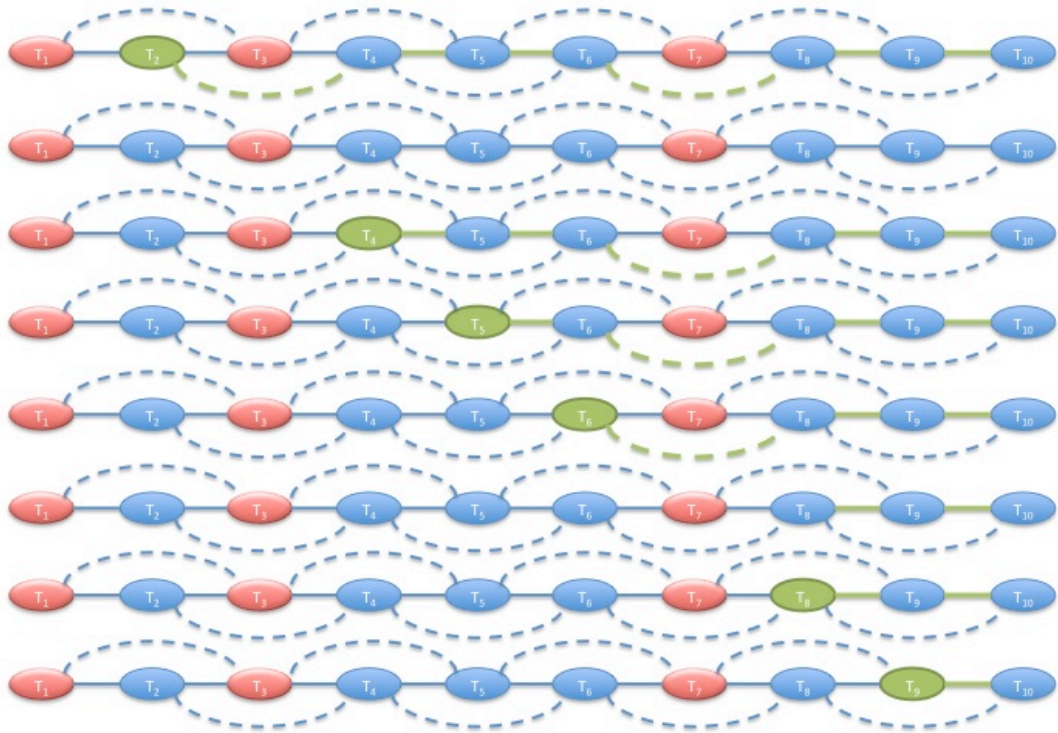

**Figure 18: Bone nodes 1, 3, and 7 have failed.**

## B. PRISM Model for Transmission Line

```
// Extended Model for Transmission Line
//
// Development Notes
//=================================
// 1. Each non-terminal bone node (T2-T9) has four states (sleep/operational/off/done), instead of just two (on/off)
//    State0: BROKEN: Can be recovered.
//            State-Reward "failuretime" applies.
//    State1: OPERATIONAL: TX takes place. RX may take place. Device can break down. Done after (a) TX transmission.
//            Transition-Rewards "sentpacketsTi" and "receivedpacketsTi" apply.
//            Transition-Reward "backupij" applies for each backup transmission.
//            Reward "batteryTi" applies for each TX/RX transmission.
//    State2: SLEEP: Can wake up. Device can break down.
//            Reward "batteryTi" applies for the time spent in this state.
//    State3: DONE: RX may take place. Device can break down.
//            Transition-Reward "receivedpacketsTi" applies.
//            Reward "batteryTi" applies for each RX transmission.
//   - duty cycle: SLEEP(2)->OPERATIONAL(1)->DONE(3)->SLEEP(2)
//   - break down: AnyMode(1,2,3) -> BROKEN(0)
//   - recovery:   BROKEN(0) -> SLEEP(2) (INSTEAD OF GOING TO OPERATIONAL AND CONSUMING ENERGY IN WAIN)
//
// 2. Each terminal bone node (T1 and T10) has three states (sleep/operational/off/), instead of just two (on/off)
//    State0: BROKEN: Can be recovered.
//            State-Reward "failuretime" applies.
//    State1: OPERATIONAL: RX takes place. Device can break down.
//            Transition-Reward "receivedpacketsTi" applies.
//            Reward "batteryTi" applies for each RX transmission.
//    State2: SLEEP: Can wake up. Device can break down.
//            Reward "batteryTi" applies for the time spent in this state.
//   - duty cycle: SLEEP(2)->OPERATIONAL(1)->LEEP(2)
//   - break down: AnyMode(1,2,3) -> BROKEN(0)
//   - recovery:   BROKEN(0) -> SLEEP(2) (INSTEAD OF GOING TO OPERATIONAL AND CONSUMING ENERGY IN WAIN)
//
// 3. In order to wake up all the bone nodes syncrhonously there is a new ENVIRONMENT module
//    All the rates are used in this module.
//    A timer can be implemented in this module in Erlang distribution to realize deterministic delays
//
// 4. Battery consumption is measured in details: depending on TX10, TX20, RX, and Sleep modes
//
// 5. A non-terminal node recovers to SLEEP state.
//    All the guards are modified to support this change.
//    A recovered node does not block sleep or wakeup.
//    Besides, sleep and wakeup should strictly alternate, which required a boolean variable in the ENVIRONMENT.

ctmc

// INSERT INPUT VALUES HERE
const int    tSLEEP=1;              // Duration of SLEEP in hours
const double tOPERATION=0.01;       // Duration of OPERATION in hours (includes LISTENING and TRANSMIT modes)
const int    tLIFE=10000;           // Average lifetime of a bone node (in hours)
const int    tRECOVERY=50;          // Average recoverytime of a bone node (in hours)
const int    tTX=5;                 // Average transmission time  (in MILLISECONDS)

const int    cTX10=120;             // Battery consumption in TX 10dBm mode
const int    cTX20=200;             // Battery consumption in TX 20dBm mode
const int    cRX=18;                // Battery consumption in RX mode
const double cSleep=0.005;          // Battery consumption in Sleep mode

global brokendevices : [0..10] init 0;

const double rSLEEP=1/tSLEEP;              // Sleep rate
const double rOPERATION=1/tOPERATION;      // Opertaion rate
```

```
const double rFAIL=1/tLIFE;              // Failure rate: once every 10000 hours
const double rRECOVERY=1/tRECOVERY;      // Recovery time:  50 hours
const double rTX=tTX*1000*60*60;         // Transmission rate

// waking up the bone nodes (stochastically)
module environment
          sleeping : bool init true;

          [wakeup]sleeping -> rSLEEP: (sleeping'=false);

          // if all the (non-broken and) sending nodes are done
          [sleep] !sleeping
                    &  (state2!=1)
                    & (state3!=1)
                    & (state4!=1)
                    & (state5!=1)
                    & (state6!=1)
                    & (state7!=1)
                    & (state8!=1)
                    & (state9!=1)
                                        -> rOPERATION:(sleeping'=true);

          // defining rate for transmissions
          // from node 2
          [TX21] true -> rTX: true;
          [TX23] true -> rTX: true;
          [TX24] true -> rTX: true;
          // from node 3
          [TX31a] true -> rTX: true;
          [TX31b] true -> rTX: true;
          [TX32] true -> rTX: true;
          [TX34a] true -> rTX: true;
          [TX34b] true -> rTX: true;
          [TX35] true -> rTX: true;
          // from node 4
          [TX42a] true -> rTX: true;
          [TX42b] true -> rTX: true;
          [TX43] true -> rTX: true;
          [TX45a] true -> rTX: true;
          [TX45b] true -> rTX: true;
          [TX46] true -> rTX: true;
          // from node 5
          [TX53a] true -> rTX: true;
          [TX53b] true -> rTX: true;
          [TX54] true -> rTX: true;
          [TX56a] true -> rTX: true;
          [TX56b] true -> rTX: true;
          [TX57] true -> rTX: true;
          // from node 6
          [TX64] true -> rTX: true;
          [TX65a] true -> rTX: true;
          [TX65b] true -> rTX: true;
          [TX67] true -> rTX: true;
          [TX68a] true -> rTX: true;
          [TX68b] true -> rTX: true;
          // from node 7
          [TX75] true -> rTX: true;
          [TX76a] true -> rTX: true;
          [TX76b] true -> rTX: true;
          [TX78] true -> rTX: true;
          [TX79a] true -> rTX: true;
          [TX79b] true -> rTX: true;
          // from node 8
          [TX86] true -> rTX: true;
          [TX87a] true -> rTX: true;
          [TX87b] true -> rTX: true;
          [TX89] true -> rTX: true;
```

```
            [TX810a] true -> rTX: true;
            [TX810b] true -> rTX: true;
            // from node 9
            [TX97] true -> rTX: true;
            [TX98] true -> rTX: true;
            [TX910] true -> rTX: true;
endmodule

module tower1
            state1 : [0..2] init 2; // 0: broken 1: operational 2: sleeping

            // wake up
            [wakeup] state1=2 -> (state1'=1);
            // broken node shouldn't block the wakeup transitions of other nodes
            [wakeup] state1=0 -> true;

            // regular receive
            [TX21] state1=1 -> true;
            // expensive receive
            [TX31a] state1=1 & (state2=0 | state2=2) & (state4=0 | state4=0)  -> true;
            [TX31b] state1=1 & (state2=0 | state2=0) & (state10=0 | state10=2)  -> true;

            // go to sleep
            [sleep] state1=1 -> (state1'=2);
            // broken node shouldn't block the sleep transitions of other nodes
            [sleep] state1=0 -> true;

            // physical failure
            [] state1>0 & brokendevices<10 -> rFAIL: (state1'=0) & (brokendevices'=brokendevices+1);
            // recovery
            [] state1=0 & brokendevices>0 -> rRECOVERY: (state1'=2) & (brokendevices'=brokendevices-1);
endmodule

module tower2
            state2 : [0..3] init 2; // 0: broken 1: operational 2: sleeping 3: done

            // wake up
            [wakeup] state2=2 -> (state2'=1);
            // broken node shouldn't block the wakeup transitions of other nodes
            [wakeup] state2=0 -> true;

            // regular receive
            [TX32] (state2=1 | state2=3) -> true;
            // expensive receive
            [TX42a] (state2=1 | state2=3) & (state3=0 | state3=2) & (state5=0 | state5=2)  -> true;
            [TX42b] (state2=1 | state2=3) & (state3=0 | state3=2) & (state10=0 | state10=2)  -> true;

            // regular transmission
            [TX21] state2=1 -> (state2'=3);
            // rerouted transmission
            [TX23] state2=1 & (state1=0 | state1=2) -> (state2'=3);
            // expensive transmission
            [TX24] state2=1 & (state1=0 | state1=2) & (state3=0 | state3=2) -> (state2'=3);

            // go to sleep
            [sleep] state2=3 -> (state2'=2);
            // broken or just recovered node shouldn't block the sleep transitions of other nodes
            [sleep] state2=0 | state2=2 -> true;

            // physical failure
            [] state2>0 & brokendevices<10 -> rFAIL: (state2'=0) & (brokendevices'=brokendevices+1);
            // recovery
            [] state2=0 & brokendevices>0 -> rRECOVERY: (state2'=2) & (brokendevices'=brokendevices-1);
endmodule

module tower3
            state3 : [0..3] init 2; //0: broken 1: operational 2: sleeping  3: done
```

```
            // wake up
            [wakeup] state3=2 -> (state3'=1);
            // broken node shouldn't block the wakeup transitions of other nodes
            [wakeup] state3=0 -> true;

            // regular receive
            [TX43] (state3=1 | state3=3) -> true;
            // rerouted receive
            [TX23] (state3=1 | state3=3) & (state1=0 | state1=2) -> true;
            // expensive receive
            [TX53a] (state3=1 | state3=3) & (state4=0 | state4=2) & (state6=0 | state6=2) -> true;
            [TX53b] (state3=1 | state3=3) & (state4=0 | state4=2) & (state10=0 | state10=2) -> true;

            // regular transmission
            [TX32] state3=1 -> (state3'=3);
            // rerouted transmission
            [TX34a] state3=1 & (state1=0 | state1=2) -> (state3'=3);
            [TX34b] state3=1 & (state2=0 | state2=0) -> (state3'=3);
            // expensive transmission
            [TX31a] state3=1 & (state2=0 | state2=0) & (state4=0 | state4=2) -> (state3'=3);
            [TX31b] state3=1 & (state2=0 | state2=0) & (state10=0 | state10=2) -> (state3'=3);
            [TX35] state3=1 & (state1=0 | state1=2) & (state4=0 | state4=2) -> (state3'=3);

            // go to sleep
            [sleep] state3=3 -> (state3'=2);
            // broken or just recovered node shouldn't block the sleep transitions of other nodes
            [sleep] state3=0 | state3=2 -> true;

            // physical failure
            [] state3>0 & brokendevices<10 -> rFAIL: (state3'=0) & (brokendevices'=brokendevices+1);
            // recovery
            [] state3=0 & brokendevices>0 -> rRECOVERY: (state3'=2) & (brokendevices'=brokendevices-1);
endmodule

module tower4
            state4 : [0..3] init 2; //0: broken 1: operational 2: sleeping 3: done

            // wake up
            [wakeup] state4=2 -> (state4'=1);
            // broken node shouldn't block the wakeup transitions of other nodes
            [wakeup] state4=0 -> true;

            // regular receive
            [TX54] (state4=1 | state4=3) -> true;
            // rerouted receive
            [TX34a] (state4=1 | state4=3) & (state1=0 | state1=2) -> true;
            [TX34b] (state4=1 | state4=3) & (state2=0 | state2=0) -> true;
            // expensive receive
            [TX24] (state4=1 | state4=3) & (state1=0 | state1=2) & (state3=0 | state3=2) -> true;
            [TX64] (state4=1 | state4=3) & (state5=0 | state5=2) & (state10=0 | state10=2) -> true;

            // regular transmission
            [TX43] state4=1 -> (state4'=3);
            // rerouted transmission
            [TX45a] state4=1 & (state1=0 | state1=2) -> (state4'=3);
            [TX45b] state4=1 & (state3=0 | state3=2) -> (state4'=3);
            // expensive transmission
            [TX42a] state4=1 & (state3=0 | state3=2) & (state5=0 | state5=2) -> (state4'=3);
            [TX42b] state4=1 & (state3=0 | state3=2) & (state10=0 | state10=2) -> (state4'=3);
            [TX46] state4=1 & (state1=0 | state1=2) & (state5=0 | state5=2) -> (state4'=3);

            // go to sleep
            [sleep] state4=3 -> (state4'=2);
            // broken or just recovered node shouldn't block the sleep transitions of other nodes
            [sleep] state4=0 | state4=2 -> true;
```

```
            // physical failure
            [] state4>0 & brokendevices<10 -> rFAIL: (state4'=0) & (brokendevices'=brokendevices+1);
            // recovery
            [] state4=0 & brokendevices>0 -> rRECOVERY: (state4'=2) & (brokendevices'=brokendevices-1);
endmodule

module tower5
            state5 : [0..3] init 2; //0: broken 1: operational  2: sleeping  3: done

            // wake up
            [wakeup] state5=2 -> (state5'=1);
            // broken node shouldn't block the wakeup transitions of other nodes
            [wakeup] state5=0 -> true;

            // rerouted receive
            [TX45a] (state5=1 | state5=3) & (state1=0 | state1=2) ->  true;
            [TX45b] (state5=1 | state5=3) & (state3=0 | state3=2) ->  true;
            [TX65a] (state5=1 | state5=3) & (state7=0 | state7=2) ->  true;
            [TX65b] (state5=1 | state5=3) & (state10=0 | state10=2) ->  true;
            // expensive receive
            [TX35] (state5=1 | state5=3) & (state1=0 | state1=2) & (state4=0 | state4=2) ->  true;
            [TX75] (state5=1 | state5=3) & (state6=0 | state6=2) & (state10=0 | state10=2) ->  true;

            // regular transmission
            [TX54] state5=1 ->  (state5'=3);
            // rerouted transmission
            [TX56a] state5=1 & (state1=0 | state1=2) -> (state5'=3);
            [TX56b] state5=1 & (state4=0 | state4=2) -> (state5'=3);
            // expensive transmission
            [TX53a] state5=1 & (state4=0 | state4=2) & (state6=0 | state6=2) ->  (state5'=3);
            [TX53b] state5=1 & (state4=0 | state4=2) & (state10=0 | state10=2) ->  (state5'=3);
            [TX57] state5=1 & (state1=0 | state1=2) & (state6=0 | state6=2) ->  (state5'=3);

            // go to sleep
            [sleep] state5=3 -> (state5'=2);
            // broken or just recovered node shouldn't block the sleep transitions of other nodes
            [sleep] state5=0 | state5=2 -> true;

            // physical failure
            [] state5>0 & brokendevices<10 -> rFAIL: (state5'=0) & (brokendevices'=brokendevices+1);
            // recovery
            [] state5=0 & brokendevices>0 -> rRECOVERY: (state5'=2) & (brokendevices'=brokendevices-1);
endmodule

module tower6
            state6 : [0..3] init 2; //0: broken 1: operational  2: sleeping  3: done

            // wake up
            [wakeup] state6=2 -> (state6'=1);
            // broken node shouldn't block the wakeup transitions of other nodes
            [wakeup] state6=0 -> true;

            // rerouted receive
            [TX56a] (state6=1 | state6=3) & (state1=0 | state1=2) ->  true;
            [TX56b] (state6=1 | state6=3) & (state4=0 | state4=2) ->  true;
            [TX76a] (state6=1 | state6=3) & (state8=0 | state8=2) ->  true;
            [TX76b] (state6=1 | state6=3) & (state10=0 | state10=2) ->  true;
            // expensive receive
            [TX46] (state6=1 | state6=3) & (state1=0 | state1=2) & (state5=0 | state5=2) ->  true;
            [TX86] (state6=1 | state6=3) & (state7=0 | state7=2) & (state10=0 | state10=2) ->  true;

            // regular transmission
            [TX67] state6=1 ->  (state6'=3);
            // rerouted transmission
            [TX65a] state6=1 & (state7=0 | state7=2) -> (state6'=3);
```

```
                [TX65b] state6=1 & (state10=0 | state10=2) ->  (state6'=3);
                // expensive transmission
                [TX64] state6=1 & (state10=0 | state10=2) & (state5=0 | state5=2) ->  (state6'=3);
                [TX68b] state6=1 & (state5=0 | state5=2) & (state7=0 | state7=2) ->  (state6'=3);
                [TX68a] state6=1 & (state1=0 | state1=2) & (state7=0 | state7=2) ->  (state6'=3);

                // go to sleep
                [sleep] state6=3 -> (state6'=2);
                // broken or just recovered node shouldn't block the sleep transitions of other nodes
                [sleep] state6=0 | state6=2 -> true;

                // physical failure
                [] state6>0 & brokendevices<10 -> rFAIL: (state6'=0) & (brokendevices'=brokendevices+1);
                // recovery
                [] state6=0 & brokendevices>0 -> rRECOVERY: (state6'=2) & (brokendevices'=brokendevices-1);
endmodule

module tower7
                state7 : [0..3] init 2; //0: broken 1: operational  2: sleeping  3: done

                // wake up
                [wakeup] state7=2 -> (state7'=1);
                // broken node shouldn't block the wakeup transitions of other nodes
                [wakeup] state7=0 -> true;

                // regular receive
                [TX67] (state7=1 | state7=3) ->  true;
                // rerouted receive
                [TX87b] (state7=1 | state7=3) & (state9=0 | state9=2) ->  true;
                [TX87a] (state7=1 | state7=3) & (state10=0 | state10=2) ->  true;
                // expensive receive
                [TX57] (state7=1 | state7=3) & (state1=0 | state1=2) & (state6=0 | state6=2) ->  true;
                [TX97] (state7=1 | state7=3) & (state8=0 | state8=2) & (state10=0 | state10=2) ->  true;

                // regular transmission
                [TX78] state7=1 ->  (state7'=3);
                // rerouted transmission
                [TX76b] state7=1 & (state10=0 | state10=2) ->  (state7'=3);
                [TX76a] state7=1 & (state8=0 | state8=2) ->  (state7'=3);
                // expensive transmission
                [TX75] state7=1 & (state10=0 | state10=2) & (state6=0 | state6=2) ->  (state7'=3);
                [TX79b] state7=1 & (state6=0 | state6=2) & (state8=0 | state8=2) ->  (state7'=3);
                [TX79a] state7=1 & (state1=0 | state1=2) & (state8=0 | state8=2) ->  (state7'=3);

                // go to sleep
                [sleep] state7=3 -> (state7'=2);
                // broken or just recovered node shouldn't block the sleep transitions of other nodes
                [sleep] state7=0 | state7=2 -> true;

                // physical failure
                [] state7>0 & brokendevices<10 -> rFAIL: (state7'=0) & (brokendevices'=brokendevices+1);
                // recovery
                [] state7=0 & brokendevices>0 -> rRECOVERY: (state7'=2) & (brokendevices'=brokendevices-1);
endmodule

module tower8
                state8 : [0..3] init 2; //0: broken 1: operational  2: sleeping  3: done

                // wake up
                [wakeup] state8=2 -> (state8'=1);
                // broken node shouldn't block the wakeup transitions of other nodes
                [wakeup] state8=0 -> true;

                // regular receive
                [TX78] (state8=1 | state8=3) ->  true;
                // rerouted receive
                [TX98] (state8=1 | state8=3) & (state10=0 | state10=2) ->  true;
```

```
        // expensive receive
        [TX68a] (state8=1 | state8=3) & (state1=0 | state1=2) & (state7=0 | state7=2) -> true;
        [TX68b] (state8=1 | state8=3) & (state5=0 | state5=2) & (state7=0 | state7=2) -> true;

        // regular transmission
        [TX89] state8=1 ->  (state8'=3);
        // rerouted transmission
        [TX87a] state8=1 & (state10=0 | state10=2) ->  (state8'=3);
        [TX87b] state8=1 & (state9=0 | state9=2) ->  (state8'=3);
        // expensive transmission
        [TX86] state8=1 & (state10=0 | state10=2) & (state7=0 | state7=2) ->  (state8'=3);
        [TX810b] state8=1 & (state7=0 | state7=2) & (state9=0 | state9=2) ->  (state8'=3);
        [TX810a] state8=1 & (state1=0 | state1=2) & (state9=0 | state9=2) ->  (state8'=3);

        // go to sleep
        [sleep] state8=3 -> (state8'=2);
        // broken or just recovered node shouldn't block the sleep transitions of other nodes
        [sleep] state8=0 | state8=2 -> true;

        // physical failure
        [] state8>0 & brokendevices<10 -> rFAIL: (state8'=0) & (brokendevices'=brokendevices+1);
        // recovery
        [] state8=0 & brokendevices>0 -> rRECOVERY: (state8'=2) & (brokendevices'=brokendevices-1);
endmodule

module tower9
        state9 : [0..3] init 2; //0: broken  1: operational  2: sleeping  3: done

        // wake up
        [wakeup] state9=2 -> (state9'=1);
        // broken node shouldn't block the wakeup transitions of other nodes
        [wakeup] state9=0 -> true;

        // regular receive
        [TX89] (state9=1 | state9=3) ->  true;
        // expensive receive
        [TX79a] (state9=1 | state9=3) & (state1=0 | state1=2) & (state8=0 | state8=2) -> true;
        [TX79b] (state9=1 | state9=3) & (state6=0 | state6=2) & (state8=0 | state8=2) -> true;

        // regular transmission
        [TX910] state9=1 ->  (state9'=3);
        // rerouted transmission
        [TX98] state9=1 & (state10=0 | state10=2) ->  (state9'=3);
        // expensive transmission
        [TX97] state9=1 & (state10=0 | state10=2) & (state8=0 | state8=2) ->  (state9'=3);

        // go to sleep
        [sleep] state9=3 -> (state9'=2);
        // broken or just recovered node shouldn't block the sleep transitions of other nodes
        [sleep] state9=0 | state9=2 -> true;

        // physical failure
        [] state9>0 & brokendevices<10 -> rFAIL: (state9'=0) & (brokendevices'=brokendevices+1);
        // recovery
        [] state9=0 & brokendevices>0 -> rRECOVERY: (state9'=2) & (brokendevices'=brokendevices-1);

endmodule

module tower10
        state10 : [0..2] init 2; //0: broken 1: operational 2: sleeping

        // wake up
        [wakeup] state10=2 -> (state10'=1);
        // broken node shouldn't block the wakeup transitions of other nodes
        [wakeup] state10=0 -> true;

        // regular receive
```

```
            [TX910] state10=1 ->  true;
            // expensive receive
            [TX810a] state10=1 & (state1=0 | state1=2) & (state9=0 | state9=2) ->  true;
            [TX810b] state10=1 & (state7=0 | state7=2) & (state9=0 | state9=2) ->  true;

            // go to sleep
            [sleep] state10=1 -> (state10'=2);
            // broken node shouldn't block the sleep transitions of other nodes
            [sleep] state10=0 | state10=2 -> true;

            // physical failure
            [] state10>0 & brokendevices<10 -> rFAIL: (state10'=0) & (brokendevices'=brokendevices+1);
            // recovery
            [] state10=0 & brokendevices>0 -> rRECOVERY: (state10'=2) & (brokendevices'=brokendevices-1);
endmodule

// USAGE OF BACKUP LINKS
//
rewards "backup13"
[TX31a] state3=1 & (state2=0 | state2=0) & (state4=0 | state4=2) :1;
[TX31b] state3=1 & (state2=0 | state2=0) & (state10=0 | state10=2) :1;
endrewards

rewards "backup24"
[TX24] state2=1 & (state1=0 | state1=2) & (state3=0 | state3=2) :1;
[TX42a] state4=1 & (state3=0 | state3=2) & (state5=0 | state5=2) :1;
[TX42b] state4=1 & (state3=0 | state3=2) & (state10=0 | state10=2) :1;
endrewards

rewards "backup35"
[TX35] state3=1 & (state1=0 | state1=2) & (state4=0 | state4=2) :1;
[TX53a] state5=1 & (state4=0 | state4=2) & (state6=0 | state6=2) :1;
[TX53b] state5=1 & (state4=0 | state4=2) & (state10=0 | state10=2) :1;
endrewards

rewards "backup46"
[TX46] state4=1 & (state1=0 | state1=2) & ((state5=0 | state5=2) | state5=2) :1;
[TX64] state6=1 & (state10=0 | state10=2) & (state5=0 | state5=2) :1;
endrewards

rewards "backup57"
[TX57] state5=1 & (state1=0 | state1=2) & (state6=0 | state6=2) :1;
[TX75] state7=1 & (state10=0 | state10=2) & (state6=0 | state6=2) :1;
endrewards

rewards "backup68"
[TX68b] state6=1 & (state5=0 | state5=2) & (state7=0 | state7=2) :1;
[TX68a] state6=1 & (state1=0 | state1=2) & (state7=0 | state7=2) :1;
[TX86] state8=1 & (state10=0 | state10=2) & (state7=0 | state7=2) :1;
endrewards

rewards "backup79"
[TX79b] state7=1 & (state6=0 | state6=2) & (state8=0 | state8=2) :1;
[TX79a] state7=1 & (state1=0 | state1=2) & (state8=0 | state8=2) :1;
[TX97] state9=1 & (state10=0 | state10=2) & (state8=0 | state8=2) :1;
endrewards

rewards "backup810"
[TX810b] state8=1 & (state7=0 | state7=2) & (state9=0 | state9=2) :1;
[TX810a] state8=1 & (state1=0 | state1=2) & (state9=0 | state9=2) :1;
endrewards

// BATTERY CONSUMPTION
//===============================
rewards "battery1"
[] state1=2 : cSleep;
[TX21] state1=1 :cRX;
```

```
[TX31a] state1=1 & (state2=0 | state2=0) & (state4=0 | state4=2)  :cRX;
[TX31b] state1=1 & (state2=0 | state2=0) & (state10=0 | state10=2)  :cRX;endrewards

rewards "battery2"
[] state2=2 : cSleep;
[TX21] state2=1 :cTX10;
[TX23] state2=1 & (state1=0 | state1=2) :cTX10;
[TX24] state2=1 & (state1=0 | state1=2) & (state3=0 | state3=2) :cTX20;
[TX32] (state2=1 | state2=3) :cRX;
[TX42a] (state2=1 | state2=3) & (state3=0 | state3=2) & (state5=0 | state5=2)  :cRX;
[TX42b] (state2=1 | state2=3) & (state3=0 | state3=2) & (state10=0 | state10=2)  :cRX;endrewards

rewards "battery3"
[] state3=2 : cSleep;
[TX32] state3=1 :cTX10;
[TX34a] state3=1 & (state1=0 | state1=2) :cTX10;
[TX34b] state3=1 & (state2=0 | state2=0) :cTX10;
[TX31a] state3=1 & (state2=0 | state2=0) & (state4=0 | state4=2) :cTX20;
[TX31b] state3=1 & (state2=0 | state2=0) & (state10=0 | state10=2) :cTX20;
[TX35] state3=1 & (state1=0 | state1=2) & (state4=0 | state4=2) :cTX20;
[TX43] (state3=1 | state3=3) :cRX;
[TX23] (state3=1 | state3=3) & (state1=0 | state1=2) :cRX;
[TX53a] (state3=1 | state3=3) & (state4=0 | state4=2) & (state6=0 | state6=2) :cRX;
[TX53b] (state3=1 | state3=3) & (state4=0 | state4=2) & (state10=0 | state10=2) :cRX;endrewards

rewards "battery4"
[] state4=2 : cSleep;
[TX43] state4=1 : cTX10;
[TX45a] state4=1 & (state1=0 | state1=2) :cTX10;
[TX45b] state4=1 & (state3=0 | state3=2) :cTX10;
[TX42a] state4=1 & (state3=0 | state3=2) & (state5=0 | state5=2) :cTX20;
[TX42b] state4=1 & (state3=0 | state3=2) & (state10=0 | state10=2) :cTX20;
[TX46] state4=1 & (state1=0 | state1=2) & (state5=0 | state5=2) :cTX20;
[TX54] (state4=1 | state4=3) :cRX;
[TX34a] (state4=1 | state4=3) & (state1=0 | state1=2) :cRX;
[TX34b] (state4=1 | state4=3) & (state2=0 | state2=0) :cRX;
[TX24] (state4=1 | state4=3) & (state1=0 | state1=2) & (state3=0 | state3=2) :cRX;
[TX64] (state4=1 | state4=3) & (state5=0 | state5=2) & (state10=0 | state10=2) :cRX;endrewards

rewards "battery5"
[] state5=2 : cSleep;
[TX54] state5=1 :cTX10;
[TX56a] state5=1 & (state1=0 | state1=2) :cTX10;
[TX56b] state5=1 & (state4=0 | state4=2) :cTX10;
[TX53a] state5=1 & (state4=0 | state4=2) & (state6=0 | state6=2) :cTX20;
[TX53b] state5=1 & (state4=0 | state4=2) & (state10=0 | state10=2) :cTX20;
[TX57] state5=1 & (state1=0 | state1=2) & (state6=0 | state6=2) :cTX20;
[TX45a] (state5=1 | state5=3) & (state1=0 | state1=2) :cRX;
[TX45b] (state5=1 | state5=3) & (state3=0 | state3=2) :cRX;
[TX65a] (state5=1 | state5=3) & (state7=0 | state7=2) :cRX;
[TX65b] (state5=1 | state5=3) & (state10=0 | state10=2) :cRX;
[TX35] (state5=1 | state5=3) & (state1=0 | state1=2) & (state4=0 | state4=2) :cRX;
[TX75] (state5=1 | state5=3) & (state6=0 | state6=2) & (state10=0 | state10=2) :cRX;endrewards

rewards "battery6"
[] state6=2 : cSleep;
[TX67] state6=1 :cTX10;
[TX65b] state6=1 & (state10=0 | state10=2) :cTX10;
[TX65a] state6=1 & (state7=0 | state7=2) :cTX10;
[TX64] state6=1 & (state10=0 | state10=2) & (state5=0 | state5=2) :cTX20;
[TX68b] state6=1 & (state5=0 | state5=2) & (state7=0 | state7=2) :cTX20;
[TX68a] state6=1 & (state1=0 | state1=2) & (state7=0 | state7=2) :cTX20;
[TX56a] (state6=1 | state6=3) & (state1=0 | state1=2) :cRX;
[TX56b] (state6=1 | state6=3) & (state4=0 | state4=2) :cRX;
[TX76a] (state6=1 | state6=3) & (state8=0 | state8=2) :cRX;
[TX76b] (state6=1 | state6=3) & (state10=0 | state10=2) :cRX;
[TX46] (state6=1 | state6=3) & (state1=0 | state1=2) & (state5=0 | state5=2) :cRX;
```

[TX86] (state6=1 | state6=3) & (state7=0 | state7=2) & (state10=0 | state10=2) :cRX;endrewards

rewards "battery7"
[] state7=2 : cSleep;
[TX78] state7=1 :cTX10;
[TX76b] state7=1 & (state10=0 | state10=2) :cTX10;
[TX76a] state7=1 & (state8=0 | state8=2) :cTX10;
[TX75] state7=1 & (state10=0 | state10=2) & (state6=0 | state6=2) :cTX20;
[TX79b] state7=1 & (state6=0 | state6=2) & (state8=0 | state8=2) :cTX20;
[TX79a] state7=1 & (state1=0 | state1=2) & (state8=0 | state8=2) :cTX20;
[TX67] (state7=1 | state7=3) :cRX;
[TX87b] (state7=1 | state7=3) & (state9=0 | state9=2) :cRX;
[TX87a] (state7=1 | state7=3) & (state10=0 | state10=2) :cRX;
[TX57] (state7=1 | state7=3) & (state1=0 | state1=2) & (state6=0 | state6=2) :cRX;
[TX97] (state7=1 | state7=3) & (state8=0 | state8=2) & (state10=0 | state10=2) :cRX;endrewards

rewards "battery8"
[] state8=2 : cSleep;
[TX89] state8=1 :cTX10;
[TX87b] state8=1 & (state10=0 | state10=2) :cTX10;
[TX87a] state8=1 & (state9=0 | state9=2) :cTX10;
[TX86] state8=1 & (state10=0 | state10=2) & (state7=0 | state7=2) :cTX20;
[TX810b] state8=1 & (state7=0 | state7=2) & (state9=0 | state9=2) :cTX20;
[TX810a] state8=1 & (state1=0 | state1=2) & (state9=0 | state9=2) :cTX20;
[TX78] (state8=1 | state8=3) :cRX;
[TX98] (state8=1 | state8=3) & (state10=0 | state10=2) :cRX;
[TX68a] (state8=1 | state8=3) & (state1=0 | state1=2) & (state7=0 | state7=2) :cRX;
[TX68b] (state8=1 | state8=3) & (state5=0 | state5=2) & (state7=0 | state7=2) :cRX;endrewards

rewards "battery9"
[] state9=2 : cSleep;
[TX910] state9=1 :cTX10;
[TX98] state9=1 & (state10=0 | state10=2) :cTX10;
[TX97] state9=1 & (state10=0 | state10=2) & (state8=0 | state8=2) :cTX20;
[TX89] (state9=1 | state9=3) :cRX;
[TX79a] (state9=1 | state9=3) & (state1=0 | state1=2) & (state8=0 | state8=2) :cRX;
[TX79b] (state9=1 | state9=3) & (state6=0 | state6=2) & (state8=0 | state8=2) :cRX;endrewards

rewards "battery10"
[] state10=2 : cSleep;
[TX910] state10=1 :cRX;
[TX810a] state10=1 & (state1=0 | state1=2) & (state9=0 | state9=2) :cRX;
[TX810b] state10=1 & (state7=0 | state7=2) & (state9=0 | state9=2) :cRX;endrewards

// NUMBER OF RECEIVED PACKETS (Received in T1 and T10)
//=================================
rewards "receivedpacketsT1" // packets received in T1
[TX21] state1=1 :1;
[TX31a] state1=1 & (state2=0 | state2=0) & (state4=0 | state4=2) :1;
[TX31b] state1=1 & (state2=0 | state2=0) & (state10=0 | state10=2) :1;endrewards

rewards "receivedpacketsT2"
[TX32] (state2=1 | state2=3) :1;
[TX42a] (state2=1 | state2=3) & (state3=0 | state3=2) & (state5=0 | state5=2) :1;
[TX42b] (state2=1 | state2=3) & (state3=0 | state3=2) & (state10=0 | state10=2) :1;endrewards

rewards "receivedpacketsT3"
[TX43] (state3=1 | state3=3) :1;
[TX23] (state3=1 | state3=3) & (state1=0 | state1=2) :1;
[TX53a] (state3=1 | state3=3) & (state4=0 | state4=2) & (state6=0 | state6=2) :1;
[TX53b] (state3=1 | state3=3) & (state4=0 | state4=2) & (state10=0 | state10=2) :1;endrewards

rewards "receivedpacketsT4"
[TX54] (state4=1 | state4=3) :1;
[TX34a] (state4=1 | state4=3) & (state1=0 | state1=2) :1;
[TX34b] (state4=1 | state4=3) & (state2=0 | state2=0) :1;
[TX24] (state4=1 | state4=3) & (state1=0 | state1=2) & (state3=0 | state3=2) :1;

[TX64] (state4=1 | state4=3) & (state5=0 | state5=2) & (state10=0 | state10=2) :1;endrewards

rewards "receivedpacketsT5"
[TX45a] (state5=1 | state5=3) & (state1=0 | state1=2) :1;
[TX45b] (state5=1 | state5=3) & (state3=0 | state3=2) :1;
[TX65a] (state5=1 | state5=3) & (state7=0 | state7=2) :1;
[TX65b] (state5=1 | state5=3) & (state10=0 | state10=2) :1;
[TX35] (state5=1 | state5=3) & (state1=0 | state1=2) & (state4=0 | state4=2) :1;
[TX75] (state5=1 | state5=3) & (state6=0 | state6=2) & (state10=0 | state10=2) :1;endrewards

rewards "receivedpacketsT6"
[TX56a] (state6=1 | state6=3) & (state1=0 | state1=2) :1;
[TX56b] (state6=1 | state6=3) & (state4=0 | state4=2) :1;
[TX76a] (state6=1 | state6=3) & (state8=0 | state8=2) :1;
[TX76b] (state6=1 | state6=3) & (state10=0 | state10=2) :1;
[TX46] (state6=1 | state6=3) & (state1=0 | state1=2) & (state5=0 | state5=2) :1;
[TX86] (state6=1 | state6=3) & (state7=0 | state7=2) & (state10=0 | state10=2) :1;endrewards

rewards "receivedpacketsT7"
[TX67] (state7=1 | state7=3) :1;
[TX87b] (state7=1 | state7=3) & (state9=0 | state9=2) :1;
[TX87a] (state7=1 | state7=3) & (state10=0 | state10=2) :1;
[TX57] (state7=1 | state7=3) & (state1=0 | state1=2) & (state6=0 | state6=2) :1;
[TX97] (state7=1 | state7=3) & (state8=0 | state8=2) & (state10=0 | state10=2) :1;endrewards

rewards "receivedpacketsT8"
[TX78] (state8=1 | state8=3) :1;
[TX98] (state8=1 | state8=3) & (state10=0 | state10=2) :1;
[TX68a] (state8=1 | state8=3) & (state1=0 | state1=2) & (state7=0 | state7=2) :1;
[TX68b] (state8=1 | state8=3) & (state5=0 | state5=2) & (state7=0 | state7=2) :1;endrewards

rewards "receivedpacketsT9"
[TX89] (state9=1 | state9=3) :1;
[TX79a] (state9=1 | state9=3) & (state1=0 | state1=2) & (state8=0 | state8=2) :1;
[TX79b] (state9=1 | state9=3) & (state6=0 | state6=2) & (state8=0 | state8=2) :1;endrewards

rewards "receivedpacketsT10"
[TX910] state10=1 :1;
[TX810a] state10=1 & (state1=0 | state1=2) & (state9=0 | state9=2) :1;
[TX810b] state10=1 & (state7=0 | state7=2) & (state9=0 | state9=2) :1;endrewards

// NUMBER OF SENT PACKETS (ORIGINATING FORM T2-T9)
//==============================
rewards "sentpacketsT2" // packets originated from T2
[TX21] state2=1 :1;
[TX23] state2=1 & (state1=0 | state1=2) :1;
[TX24] state2=1 & (state1=0 | state1=2) & (state3=0 | state3=2) :1;endrewards

rewards "sentpacketsT3" // packets originated from T3
[TX32] state3=1 :1;
[TX34a] state3=1 & (state1=0 | state1=2) :1;
[TX34b] state3=1 & (state2=0 | state2=0) :1;
[TX31a] state3=1 & (state2=0 | state2=0) & (state4=0 | state4=2) :1;
[TX31b] state3=1 & (state2=0 | state2=0) & (state10=0 | state10=2) :1;
[TX35] state3=1 & (state1=0 | state1=2) & (state4=0 | state4=2) :1;endrewards

rewards "sentpacketsT4" // packets originated from T4
[TX43] state4=1 : 1;
[TX45a] state4=1 & (state1=0 | state1=2) :1;
[TX45b] state4=1 & (state3=0 | state3=2) :1;
[TX42a] state4=1 & (state3=0 | state3=2) & (state5=0 | state5=2) :1;
[TX42b] state4=1 & (state3=0 | state3=2) & (state10=0 | state10=2) :1;
[TX46] state4=1 & (state1=0 | state1=2) & (state5=0 | state5=2) :1;endrewards

rewards "sentpacketsT5" // packets originated from T5
[TX54] state5=1 :1;
[TX56a] state5=1 & (state1=0 | state1=2) :1;

[TX56b] state5=1 & (state4=0 | state4=2) :1;
[TX53a] state5=1 & (state4=0 | state4=2) & (state6=0 | state6=2) :1;
[TX53b] state5=1 & (state4=0 | state4=2) & (state10=0 | state10=2) :1;
[TX57] state5=1 & (state1=0 | state1=2) & (state6=0 | state6=2) :1;endrewards

rewards "sentpacketsT6" // packets originated from T6
[TX67] state6=1 :1;
[TX65b] state6=1 & (state10=0 | state10=2) :1;
[TX65a] state6=1 & (state7=0 | state7=2) :1;
[TX64] state6=1 & (state10=0 | state10=2) & (state5=0 | state5=2) :1;
[TX68b] state6=1 & (state5=0 | state5=2) & (state7=0 | state7=2) :1;
[TX68a] state6=1 & (state1=0 | state1=2) & (state7=0 | state7=2) :1;endrewards

rewards "sentpacketsT7" // packets originated from T7
[TX78] state7=1 :1;
[TX76b] state7=1 & (state10=0 | state10=2) :1;
[TX76a] state7=1 & (state8=0 | state8=2) :1;
[TX75] state7=1 & (state10=0 | state10=2) & (state6=0 | state6=2) :1;
[TX79b] state7=1 & (state6=0 | state6=2) & (state8=0 | state8=2) :1;
[TX79a] state7=1 & (state1=0 | state1=2) & (state8=0 | state8=2) :1;endrewards

rewards "sentpacketsT8" // packets originated from T8
[TX89] state8=1 :1;
[TX87a] state8=1 & (state10=0 | state10=2) :1;
[TX87b] state8=1 & (state9=0 | state9=2) :1;
[TX86] state8=1 & (state10=0 | state10=2) & (state7=0 | state7=2) :1;
[TX810b] state8=1 & (state7=0 | state7=2) & (state9=0 | state9=2) :1;
[TX810a] state8=1 & (state1=0 | state1=2) & (state9=0 | state9=2) :1;endrewards

rewards "sentpacketsT9" // packets originated from T9
[TX910] state9=1 :1;
[TX98] state9=1 & (state10=0 | state10=2) :1;
[TX97] state9=1 & (state10=0 | state10=2) & (state8=0 | state8=2) :1;endrewards

// FAILURE TIMES (DURATION)
//================================
rewards "fail1"
state1=0 : 1;endrewards

rewards "fail2"
state2=0 : 1;endrewards

rewards "fail3"
state3=0 : 1;endrewards

rewards "fail4"
state4=0 : 1;endrewards

rewards "fail5"
state5=0 : 1;endrewards

rewards "fail6"
state6=0 : 1;endrewards

rewards "fail7"
state7=0 : 1;endrewards

rewards "fail8"
state8=0 : 1;endrewards

rewards "fail9"
state9=0 : 1;endrewards

rewards "fail10"
state10=0 : 1;endrewards

# C. PRISM Model for Electricity Tower

// Electricity tower network, a WSN with a bone node and max 50 connected sensors in a star topology

```
// ASSUMPTIONS
//-------------------
// 0. TIME UNIT: Hour
// 1. A sensor fails with rate rFail
// 2. A failed sensor is recovered with rate rRecover

ctmc

const double rFail=0.000001;      // Failure rate per sensor
const double rRecover=0.01;       // Recovery rate per sensor
const double rSend=1;             // Packet rate per sensor (one packet per hour)

const int cSend=1;                // Cost of Send
const int MAXfailure=10;          // to be used in failcount

module failcount
            failure : [0..MAXfailure] init 0;

[fail1] failure<MAXfailure -> (failure'=failure+1);
[fail2] failure<MAXfailure -> (failure'=failure+1);
[fail3] failure<MAXfailure -> (failure'=failure+1);
[fail4] failure<MAXfailure -> (failure'=failure+1);
[fail5] failure<MAXfailure -> (failure'=failure+1);
[fail6] failure<MAXfailure -> (failure'=failure+1);
[fail7] failure<MAXfailure -> (failure'=failure+1);
[fail8] failure<MAXfailure -> (failure'=failure+1);
[fail9] failure<MAXfailure -> (failure'=failure+1);
[fail10] failure<MAXfailure -> (failure'=failure+1);

[rec1] failure>0 -> (failure'=failure-1);
[rec2] failure>0 -> (failure'=failure-1);
[rec3] failure>0 -> (failure'=failure-1);
[rec4] failure>0 -> (failure'=failure-1);
[rec5] failure>0 -> (failure'=failure-1);
[rec6] failure>0 -> (failure'=failure-1);
[rec7] failure>0 -> (failure'=failure-1);
[rec8] failure>0 -> (failure'=failure-1);
[rec9] failure>0 -> (failure'=failure-1);
[rec10] failure>0 -> (failure'=failure-1);

endmodule

module tower

// Defining sensors
s1 : bool init true;
s2 : bool init true;
s3 : bool init true;
s4 : bool init true;
s5 : bool init true;
s6 : bool init true;
s7 : bool init true;
s8 : bool init true;
s9 : bool init true;
s10 : bool init true;
//s11 : bool init true;
//s12 : bool init true;
//s13 : bool init true;
//s14 : bool init true;
//s15 : bool init true;
```

```
//s16 : bool init true;
//s17 : bool init true;
//s18 : bool init true;
//s19 : bool init true;
//s20 : bool init true;
//s21 : bool init true;
//s22 : bool init true;
//s23 : bool init true;
//s24 : bool init true;
//s25 : bool init true;
//s26 : bool init true;
//s27 : bool init true;
//s28 : bool init true;
//s29 : bool init true;
//s30 : bool init true;

// A sensor fails
[fail1] s1 -> rFail: (s1'=false);
[fail2] s2 -> rFail: (s2'=false) ;
[fail3] s3 -> rFail: (s3'=false) ;
[fail4] s4 -> rFail: (s4'=false) ;
[fail5] s5 -> rFail: (s5'=false) ;
[fail6] s6 -> rFail: (s6'=false) ;
[fail7] s7 -> rFail: (s7'=false) ;
[fail8] s8 -> rFail: (s8'=false) ;
[fail9] s9 -> rFail: (s9'=false) ;
[fail10] s10 -> rFail: (s10'=false) ;
//[] s11 & NumOfFailedSensors<MaxNumOfSensors & TotalNumberOfSensorFailures<MAX ->
//        rFail: (s11'=false) & (NumOfFailedSensors'=NumOfFailedSensors+1) & (TotalNumberOfSensorFailures'=TotalNumberOfSensorFailures+1);
//[] s12 & NumOfFailedSensors<MaxNumOfSensors & TotalNumberOfSensorFailures<MAX ->
//        rFail: (s12'=false) & (NumOfFailedSensors'=NumOfFailedSensors+1) & (TotalNumberOfSensorFailures'=TotalNumberOfSensorFailures+1);
//[] s13 & NumOfFailedSensors<MaxNumOfSensors & TotalNumberOfSensorFailures<MAX ->
//        rFail: (s13'=false) & (NumOfFailedSensors'=NumOfFailedSensors+1) & (TotalNumberOfSensorFailures'=TotalNumberOfSensorFailures+1);
//[] s14 & NumOfFailedSensors<MaxNumOfSensors & TotalNumberOfSensorFailures<MAX ->
//        rFail: (s14'=false) & (NumOfFailedSensors'=NumOfFailedSensors+1) & (TotalNumberOfSensorFailures'=TotalNumberOfSensorFailures+1);
//[] s15 & NumOfFailedSensors<MaxNumOfSensors & TotalNumberOfSensorFailures<MAX ->
//        rFail: (s15'=false) & (NumOfFailedSensors'=NumOfFailedSensors+1) & (TotalNumberOfSensorFailures'=TotalNumberOfSensorFailures+1);
//[] s16 -> rFail: (s16'=false);
//[] s17 -> rFail: (s17'=false);
//[] s18 -> rFail: (s18'=false);
//[] s19 -> rFail: (s19'=false);
//[] s20 -> rFail: (s20'=false);
//[] s21 -> rFail: (s21'=false);
//[] s22 -> rFail: (s22'=false);
//[] s23 -> rFail: (s23'=false);
//[] s24 -> rFail: (s24'=false);
//[] s25 -> rFail: (s25'=false);
//[] s26 -> rFail: (s26'=false);
//[] s27 -> rFail: (s27'=false);
//[] s28 -> rFail: (s28'=false);
//[] s29 -> rFail: (s29'=false);
//[] s30 -> rFail: (s30'=false);

// A failed sensor is recovered
[rec1] !s1 -> rRecover: (s1'=true);
[rec2] !s2 -> rRecover: (s2'=true);
[rec3] !s3 -> rRecover: (s3'=true);
[rec4] !s4 -> rRecover: (s4'=true);
[rec5] !s5 -> rRecover: (s5'=true);
[rec6] !s6 -> rRecover: (s6'=true);
[rec7] !s7 -> rRecover: (s7'=true);
[rec8] !s8 -> rRecover: (s8'=true);
```

```
[rec9] !s9 -> rRecover: (s9'=true);
[rec10] !s10 -> rRecover: (s10'=true);
//[] !s11 & NumOfFailedSensors>0 -> rRecover: (s11'=true) & (NumOfFailedSensors'=NumOfFailedSensors-1);
//[] !s12 & NumOfFailedSensors>0 -> rRecover: (s12'=true) & (NumOfFailedSensors'=NumOfFailedSensors-1);
//[] !s13 & NumOfFailedSensors>0 -> rRecover: (s13'=true) & (NumOfFailedSensors'=NumOfFailedSensors-1);
//[] !s14 & NumOfFailedSensors>0 -> rRecover: (s14'=true) & (NumOfFailedSensors'=NumOfFailedSensors-1);
//[] !s15 & NumOfFailedSensors>0 -> rRecover: (s15'=true) & (NumOfFailedSensors'=NumOfFailedSensors-1);
//[] !s16 -> rRecover: (s16'=true);s
//[] !s17 -> rRecover: (s17'=true);
//[] !s18 -> rRecover: (s18'=true);
//[] !s19 -> rRecover: (s19'=true);
//[] !s20 -> rRecover: (s20'=true);
//[] !s21 -> rRecover: (s21'=true);
//[] !s22 -> rRecover: (s22'=true);
//[] !s23 -> rRecover: (s23'=true);
//[] !s24 -> rRecover: (s24'=true);
//[] !s25 -> rRecover: (s25'=true);
//[] !s26 -> rRecover: (s26'=true);
//[] !s27 -> rRecover: (s27'=true);
//[] !s28 -> rRecover: (s28'=true);
//[] !s29 -> rRecover: (s29'=true);
//[] !s30 -> rRecover: (s30'=true);

// A sensor sends packet
[send1] s1 -> rSend: true;
[send2] s2 -> rSend: true;
[send3] s3 -> rSend: true;
[send4] s4 -> rSend: true;
[send5] s5 -> rSend: true;
[send6] s6 -> rSend: true;
[send7] s7 -> rSend: true;
[send8] s8 -> rSend: true;
[send9] s9 -> rSend: true;
[send10] s10 -> rSend: true;
//[s11] s11 -> rSend: true;
//[s12] s12 -> rSend: true;
//[s13] s13 -> rSend: true;
//[s14] s14 -> rSend: true;
//[s15] s15 -> rSend: true;
//[s16] s16 -> rSend: true;
//[s17] s17 -> rSend: true;
//[s18] s18 -> rSend: true;
//[s19] s19 -> rSend: true;
//[s20] s20 -> rSend: true;
//[s21] s21 -> rSend: true;
//[s22] s22 -> rSend: true;
//[s23] s23 -> rSend: true;
//[s24] s24 -> rSend: true;
//[s25] s25 -> rSend: true;
//[s26] s26 -> rSend: true;
//[s27] s27 -> rSend: true;
//[s28] s28 -> rSend: true;
//[s29] s29 -> rSend: true;
//[s30] s30 -> rSend: true;

endmodule

rewards "Doublefailure"
failure=2 : 1;
endrewards

rewards "Singlefailure"
failure=1 : 1;
endrewards

// Total number of packets that the bone node received
rewards "TotalNumberOfCommunicationsToBN"
```

```
            [send1] true : 1;
            [send2] true : 1;
            [send3] true : 1;
            [send4] true : 1;
            [send5] true : 1;
            [send6] true : 1;
            [send7] true : 1;
            [send8] true : 1;
            [send9] true : 1;
            [send10] true : 1;
endrewards

rewards "TotalNumberOfSensorsFailures"
            [fail1] true : 1;
            [fail2] true : 1;
            [fail3] true : 1;
            [fail4] true : 1;
            [fail5] true : 1;
            [fail6] true : 1;
            [fail7] true : 1;
            [fail8] true : 1;
            [fail9] true : 1;
            [fail10] true : 1;
endrewards

rewards "TotalNumberOfRecoveries"
            [rec1] true : 1;
            [rec2] true : 1;
            [rec3] true : 1;
            [rec4] true : 1;
            [rec5] true : 1;
            [rec6] true : 1;
            [rec7] true : 1;
            [rec8] true : 1;
            [rec9] true : 1;
            [rec10] true : 1;
endrewards

// Energy consumption of each sensor
rewards "s1"
            [send1] true : cSend;
endrewards
rewards "s2"
            [send2] true : cSend;
endrewards
rewards "s3"
            [send3] true : cSend;
endrewards
rewards "s4"
            [send4] true : cSend;
endrewards
rewards "s5"
            [send5] true : cSend;
endrewards
rewards "s6"
            [send6] true : cSend;
endrewards
rewards "s7"
            [send7] true : cSend;
endrewards
rewards "s8"
            [send8] true : cSend;
endrewards
rewards "s9"
            [send9] true : cSend;
endrewards
rewards "s10"
```

```
        [send10] true : cSend;
endrewards
```

## D. Compact PRISM Model for the Whole System

```
// Modelling Electricity Transmission Line composed of SIZE_BN number of towers and each tower has SIZE_SN number of sensors.
// time unit: 1 hour

ctmc
// TO BE FILLED IN BY THE USER
// 1. Number of nodes
const int SIZE_BN=100;                             // Number of bone nodes
const int MAX_BN_FAIL=5;                           // MAX number of bone node failures that can be tolerated
const int SIZE_SN=50;                              // Number of sensor nodes in each tower
const int MAX_SN_FAIL=50;                          // MAX number of sensor node failures that can be tolerated
// 2. Durations
const int SLEEPTIME=1;                             // Average sleeping duration for each node (for example: 1 hours)
const int MEANTIMEBETWEENFAILURE_SN=24000;         // Average lifetime without failure per sensor node (for example: 24K=1000
days)
const int MEANTIMEBETWEENFAILURE_BN=36000;         // Average lifetime without failure per bone node (for example: 36K=1500
days)
const int RECOVERYTIME_SN=48;                      // Average recovery time for sensor nodes(for example: 48 hours)
const int RECOVERYTIME_BN=36;                      // Average recovery time for bone nodes (for example: 36 hours)
const double PROCESSTIME=0.001;                    // Average processing duration (for example: 3.6 seconds)
// 3. Probabilities
const double pCHEAPLINK=0.95;                      // Probability of using the inexpensive transmission links
// 4. Costs
const int cCHEAPTX=24;                             // Cost of inter-tower + intra-tower comm. when using an inexpensive
transmission link
const int cEXPENSIVETX=40;                         // Cost of inter-tower + intra-tower comm. when using an expensive
transmission link
const int cSNTX=8;                                 // Cost of intra-tower comm. for a sensor node
const double cSLEEP_BN=0.001;                      // Cost of sleep for a bone node
const double cSLEEP_SN=0.001;                      // Cost of sleep for a sensor node
const int cPROCESS_BN=5;                           // Cost of processing for a bone node
const int cPROCESS_SN=2;                           // Cost of processing for a sensor node

// COMPUTED BY THE MODEL
const double rSLEEP=1/SLEEPTIME;
const double rFAIL_SN=1/MEANTIMEBETWEENFAILURE_SN;
const double rFAIL_BN=1/MEANTIMEBETWEENFAILURE_BN;
const double rRECOVERY_SN=1/RECOVERYTIME_SN;
const double rRECOVERY_BN=1/RECOVERYTIME_BN;
const double rPROCESS=1/PROCESSTIME;

// FORMULAE
formula osnf = 1-(0.01*(failedSN/(SIZE_SN*SIZE_BN))); // Operational Sensor Node Factor
formula obnf = 1-(0.01*(failedBN/SIZE_BN)); // Operational Sensor Node Factor

// Modelling node duty cycle
module controller
        mode : [1..2] init 1; // Modes for  nodes:
                                        // 1: Sleep
                                        // 2: Process
                // WAKEUP
                [wakeup] mode=1 & failedBN<MAX_BN_FAIL & failedSN<MAX_SN_FAIL -> rSLEEP: (mode'=2);

                // SENSE-AND-SEND OR RECEIVE-AND-ROUTE
                [process] mode=2 & failedBN<MAX_BN_FAIL & failedSN<MAX_SN_FAIL -> rPROCESS: (mode'=1);
endmodule

// Modelling broken sensor nodes
module sensorNodes
```

```
            failedSN : [0..MAX_SN_FAIL] init 0; // Number of failed sensor in this group

            // FAILURE
            [failSN] failedSN<MAX_SN_FAIL -> osnf*rFAIL_SN: (failedSN'=failedSN+1);
            // RECOVERY
            [repairSN] failedSN>0 -> rRECOVERY_SN: (failedSN'=failedSN-1);

endmodule

// Modelling broken sensor nodes
module boneNodes

            failedBN : [0..MAX_BN_FAIL] init 0; // Number of failed sensor in this group

            // FAILURE
            [failBN] failedBN<MAX_BN_FAIL -> obnf*rFAIL_BN: (failedBN'=failedBN+1);
            // RECOVERY
            [repairBN] failedBN>0 -> rRECOVERY_BN: (failedBN'=failedBN-1);

endmodule

// Modelling repair service. Bone nodes have higher priority
module repairService
            [repairBN] failedBN>0 ->  true;
            [repairSN] failedSN>0 & failedBN=0 ->  true;
endmodule

// Average energy consumption of a bone node
rewards "AvgEnergyBN"
[wakeup] true : pCHEAPLINK*cCHEAPTX + (1-pCHEAPLINK)*cEXPENSIVETX;
mode=1 : cSLEEP_BN;
mode=2 : cPROCESS_BN;
endrewards

// Average energy consumption of a sensor node
rewards "AvgEnergySN"
[wakeup] true : cSNTX;
mode=1 : cSLEEP_SN;
mode=2 : cPROCESS_SN;
endrewards
```